\documentclass[12pt]{iopart}
\usepackage{iopams}
\usepackage{hyperref}
\usepackage{bm}
\usepackage{graphicx}
\usepackage{verbatim}
\usepackage{xcolor}
\usepackage[T1]{fontenc}

\usepackage[mathscr]{eucal}
\usepackage{latexsym}
\usepackage{stmaryrd}
\usepackage{listings}
\usepackage[mathscr]{eucal}
\usepackage{latexsym}
\usepackage{stmaryrd}
\RequirePackage{geometry}
\RequirePackage{fancybox}
\RequirePackage{pstricks,soul,colortbl,color}
\RequirePackage{fancyhdr}
\RequirePackage{graphicx}
\usepackage{texdraw}
\usepackage{mathrsfs}
\RequirePackage{pstricks,soul,colortbl,color}
\usepackage{times}
\usepackage{cite}
\RequirePackage{amsfonts,amsthm,dsfont,latexsym,amssymb}
\usepackage{amssymb, amsthm, amsbsy, amsfonts, amsthm, paralist}
\usepackage{bm}
\usepackage{url}

\newcommand{\scri}{\mathscr{I}}
\begin{document}

\title{Seeking for toroidal event horizons from initially
stationary BH configurations}
\author{Marcelo Ponce $^\dag$} \ead{mponce@astro.rit.edu}
\author{Carlos Lousto $^\ddag$} \ead{lousto@astro.rit.edu}
\author{Yosef Zlochower $^\ddag$} \ead{yosef@astro.rit.edu}

\address{$^\dag$ Center for Computational Relativity and Gravitation,
  Rochester Institute of Technology, 85 Lomb Memorial Drive,
  Rochester, New York 14623, USA.}
\address{$^\ddag$ Center for Computational Relativity and Gravitation and School of Mathematical Sciences,
  Rochester Institute of Technology, 85 Lomb Memorial Drive,
  Rochester, New York 14623, USA.}

\date{\today}

\begin{abstract}
We construct and evolve non-rotating vacuum initial data with a ring
singularity, based on a simple extension of the standard
Brill-Lindquist multiple black-hole initial data, and search for event
horizons with spatial slices that are toroidal when the ring
radius is sufficiently large. While evolutions of the ring singularity
are not numerically feasible for large radii, we find some evidence, based
on configurations of multiple BHs arranged in a ring, that this
configuration leads to singular limit where the horizon width has zero
size, possibly indicating the presence of a naked singularity, when
the radius of the ring is sufficiently large. This is in agreement
with previous studies that have found that there is no apparent
horizon surrounding the ring singularity when the ring's radius is
larger than about twice its mass.
\end{abstract}

\submitto{\CQG}
\maketitle

\section{Introduction}
The recent dramatic breakthroughs in the numerical techniques to
evolve black-hole-binary (BHB) spacetimes~\cite{Pretorius:2005gq,
Campanelli:2005dd, Baker:2005vv} have led to rapid advancements in our
understanding of black-hole physics.  Notable among these advancements
are developments in mathematical relativity, including systems of PDEs
and gauge choices~\cite{Lindblom:2005qh, Gundlach:2006tw, vanMeter:2006vi}, the
exploration of the validity of the cosmic censorship
conjecture~\cite{Campanelli:2006uy,  Campanelli:2006fg,
Campanelli:2006fy, Rezzolla:2007rd, Sperhake:2007gu, Dain:2008ck}, and the
application of isolated horizon (IH)
formulae~\cite{Dreyer:2002mx,Schnetter:2006yt,Campanelli:2006fg,
Campanelli:2006fy,Cook:2007wr,Krishnan:2007pu}.
Recent studies include the algebraic classification of spacetime
post-merger of BHBs~\cite{Campanelli:2008dv ,Owen:2010vw},
 investigations of the orbital mechanics of
spinning BHBs~\cite{Campanelli:2006uy, Campanelli:2006fg,
Campanelli:2006fy, Herrmann:2007ex, Marronetti:2007ya,
Marronetti:2007wz, Berti:2007fi}, studies of the recoil from the
merger of unequal mass BHBs~\cite{Herrmann:2006ks,
Baker:2006vn, Gonzalez:2006md}, the remarkable discovery of
unexpectedly large recoil velocities from the merger of certain
spinning BHBs~\cite{Herrmann:2007ac, Campanelli:2007ew,
Campanelli:2007cga, Lousto:2008dn, Pollney:2007ss, Gonzalez:2007hi,
Brugmann:2007zj, Choi:2007eu, Baker:2007gi, Schnittman:2007ij,
Baker:2008md, Healy:2008js, Herrmann:2007zz, Herrmann:2007ex,
Tichy:2007hk, Koppitz:2007ev, Miller:2008en}, investigations into the
mapping between the BHB initial conditions (individual
masses and spins) and the final state of the merged black
hole~\cite{Boyle:2007sz, Boyle:2007ru, Buonanno:2007sv, Tichy:2008du,
Kesden:2008ga, Barausse:2009uz, Rezzolla:2008sd, Lousto:2009mf}, and
improvements in our understanding of the validity of approximate
BHB orbital calculations using post-Newtonian (PN)
methods~\cite{Buonanno:2006ui, Baker:2006ha, Pan:2007nw,
Buonanno:2007pf, Hannam:2007ik, Hannam:2007wf, Gopakumar:2007vh,
Hinder:2008kv}. There have also been notable advances in our
understanding of the small mass ratio limit, as well as hybrid
perturbative/numerical methods for evolving small mass ratio
BHBs~\cite{Gonzalez:2008bi, Lousto:2008dn, Lousto:2010tb}.

Even before the breakthrough there were important studies of the
structure of event horizons for non-stationary spacetimes.  An event
horizons is a 3D null hypersurface in spacetime that forms the
boundary of the region of causally connected to $\scri^+$.  The event
horizon is actually that part of the null surface that is caustic free
in the future.  If two null generators cross, then prior to the
crossing, the generators are not on the horizon, and a causal curve
intersecting these generators can terminate on $\scri^+$. To
understand why this is, we note that the region of the null
hypersurface containing this crossing is equivalent to a lightcone in
the local Minkowski frame. Unlike the region of the interior of the
light cone in the future of the intersection point, the region in
the past of the intersection point of the light cone is
causally connected to the region outside the light cone. In the case
of the event horizon, points on the null generators prior to a caustic
are causally connected to $\scri^+$.

 In Ref.~\cite{Friedman:1993ty} it was proven that, for asymptotically
flat spacetimes satisfying the null energy condition, all causal
(timelike or null) curves from $\scri^-$ to $\scri^+$ are deformable
to a topologically trivial curve. An important consequence of this
``topological censorship'' theorem is that constant time slices of an
event horizon must (at least in the distant future) be topologically
spherical.  Hawking proved~\cite{Hawking:1971vc, Hawking73a}
(see~\cite{Galloway:2005mf, Galloway:2006ws, Racz:2008tf} for
generalizations of this theorem to higher dimensions) that spacelike
slices of event horizons in asymptotically flat stationary spacetimes
obeying the dominant energy condition have topology $S^2$. For
non-stationary spacetimes, toroidal horizons are allowed, but the {\it
holes} in these horizons would close up fast enough to prevent causal
curves from traversing the holes.  Interestingly, the existence of a
single toroidal horizon slice implies that there is a 1-parameter
family of toroidal horizon slices in the neighborhood of this
particular horizon slice~\cite{Galloway:2006ws}.

The first numerical studies of the event horizon topologies for
non-stationary spacetimes involved the axisymmetric collapse of a
rotating toroidal distribution of dust~\cite{Hughes:1994ea,
Shapiro:1995rr} and theoretical studies of possible horizon topologies
based ellipsoidal wavefronts in Minkowski space~\cite{Husa:1999nm}.
However, as of yet, there have been no simulations published that have
unequivocally shown event horizons with toroidal topologies from the
mergers of multiple black holes (BHs) (see \cite{Diener:2003jc} for a
possible example), and very few that have shown event horizons of any
form (see~\cite{Diener:2003jc, Cohen:2008wa}).  The main reason for
this is that an event horizon is a global structure whose location is
determined by the entire future of the spacetime (in practice, one
only needs to evolve to the point where the final remnant BH
equilibrates). Thus, in order to find an event horizon, one must first
evolve the spacetime, obtain the full four-metric at all times in the
future of the initial hypersurface, and then evolve the null
generators of the last common apparent horizon (AH) backwards in time.
This presents a significant storage challenge, and since much of the
information about BHBs can be obtained from the AHs (using, for
example, the IH formalism), such an event horizon search is seldom
performed. In a typical BHB simulation, one is interested in the
masses and spins of the individual BHs when they are far away and in
the mass and spin of the remnant BH after it has equilibrated. In
these two regimes, the IH formalism provides accurate measurements of
the mass and spin of the BHs. However, near merger, where the AH and
event horizon can differ significantly,  the IH formalism will not
produce accurate results.  Moreover, the structure of the AHs will not
match that of the event horizon.  Thus, it is important to examine the
event horizon structure in the vicinity of the merger in order to gain
an understanding of how event horizons behave in highly dynamical
spacetimes.

While topological censorship forbids an event horizon from remaining
toroidal, it is interesting to see if purely vacuum configurations can
have instantaneous toroidal slices. To partially address this
question, we examine the dynamics of a spacetime with a ring-like,
rather than pointlike, singularity. This configuration was first
studied in~\cite{PhysRevD.41.1867}, where it was found that an AH does not exist if
the rings radius is sufficiently large, leading to the conjecture
that this is a naked singularity (see also~\cite{Jaramillo:2010tc}). We find evidence to support
this conjecture.  Here we extend the analysis and show evidence that
there is no common nonsingular event horizon for sufficiently large
ring radii.

While our proposed configuration would not violate cosmic censorship
because the singularity does not develop in the future of a
non-singular slice (i.e.\ all slices contain a singularity), questions
concerning the validity of the cosmic censorship are also quite
interesting.  The authors of Ref.  \cite{Shapiro:1991zza,
PhysRevD.45.2006} evolved a
prolate spheroidal distribution of collisionless gas and found that
generically, for large enough spheroids, a singular spindle forms on
the long axis of the spheroid which is apparently naked, violating
cosmic censorship. In a follow-up work in Ref.~\cite{Abrahams:1992ru}
they found vacuum configurations of linear black-hole distribution
like the ones we use here also contain naked singularities for
sufficiently long lines.  In~\cite{Lehner:2010pn}, the authors found
that naked singularities can form in 5-dimensional black-string
configurations.

In this paper we use the following nomenclature. While the event
horizon is a global 3-d hypersurface in a 4-d space, we are interested
in spacelike slices of the horizon. That is if $\Sigma_t$ is a
one-parameter family of spacelike slices that foliates the space time
and ${\cal H}$ is the event horizon, then we are interested in the
object ${\cal H}_t = \Sigma_t \cap {\cal H}$ (which may be a
disconnected set). In the sections below we  refer to these spatial
slices of the event horizon (${\cal H}_t$) as ``EHs''. 
We note that for well separated
BHs these ``EHs'' will also be AHs. In order to locate these distinct
EHs we track the null generators using the EHFinder Cactus
thorn~\cite{Diener:2003jc}, backwards in time, dropping those
generators that have crossed, leaving only those generators still on
the EH.

This paper is organized as follows. In Sec.~\ref{sec:Tor_Ring} we
discuss the ring configuration and toroidal event horizons. In
Sec.~\ref{sec:initial_data} we discuss the initial data for the
numerical simulations. In Sec.~\ref{sec:num} we discuss the numerical
techniques. In Sec.~\ref{sec:Results} we discuss the EHs found for the
discrete ring and discrete line cases, as well as the AHs found for
the continuum ring. Finally, in Sec.~\ref{sec:Discussion} we discuss
the implication of our numerical results and speculate about the
nature of the EH for the continuum ring.

\section{Toroidal Event Horizons and Black Hole Rings}
\label{sec:Tor_Ring}

While topological censorship requires that the 3-d hypersurface
corresponding to the event horizon is simply connected, there is no
such restriction on 2-dimensional spatial slices of the event horizon.
The question we wish to address in this paper is, can a configuration
of initially stationary (in the sense of having zero momentum, the
spacetime itself is not stationary and does not possess a timelike
Killing vector field) nonspinning black holes form a horizon with
topologically toroidal slices, and if so, can we find these slices
numerically?  In order to investigate these questions, we examine
configurations of black holes (BHs) arranged in a ring configuration,
as well configurations with BHs arranged in a line.  The idea behind
using a ring is that if neighboring BHs are close enough together, and
the ring is wide enough, then a common event horizon (if it exists)
should have a toroidal topology (on the 3-dimensional slice). We can
imagine constructing such a ring by keeping the total mass fixed,
while increasing the number of BHs in the ring.  We might expect
that, for a sufficiently large number of BHs, a common event horizon
will form. However, as we conjecture below based on our numerical
simulations, this event horizon may only forms in the limiting case
where there is an infinite number of BHs with infinitesimal masses.
The EH, while toroidal, may actually have zero width. Interestingly,
we can construct initial data corresponding to this limiting
distribution of BHs using the techniques of electrostatics. This
configuration was studied in Ref.~\cite{PhysRevD.41.1867}, as
well as~\cite{Jaramillo:2010tc}, where it
was shown that a common apparent horizon (AH) does not exist if the
ratio of the ring radius to mass is larger than $R/M\sim
\frac{20}{3\pi} \approx 2.12$. 

 An AH must be simply
connected~\cite{Galloway:2006ws}, hence the absence of an AH indicates
that the EH may not be simply connected. We note that it is possible,
to construct unusual slicings where the EH is spherical but an AH does
not exist. However, our initial data are based on a superposition of
Brill-Lindquist BHs~\cite{Brill63}, which, at least for finite numbers
of BHs, does not lead to these unusual slices.
In~\cite{PhysRevD.41.1867} it was argued that the absence on an AH
indicates that this singularity is naked.

\subsection{Initial Data}
    \label{sec:initial_data}
We construct initial data by superimposing conformally-flat, initially
stationary non-spinning, black-hole (BH) configurations.  That is, we
take as initial data $K_{ij}=0$ and $\gamma_{ij} = \psi^4
\delta_{ij}$, where $\Delta \psi = 0$. For the case of discrete BHs,
we use an ordinary superposition of Brill-Lindquist
BHs~\cite{Brill63}, while for the case of the BH ring, we use the
techniques of electrostatic to solve for the potential of a
1-dimensional ring of charge.

{\it Continuous Ring}:
To construct initial data for the continuous BH ring we solve
\begin{equation}
    \label{eq:ring_sing--sph_coord}
    \Psi(\vec{r}) = M \int_0^{2\pi}
         \frac{d\phi'}{\sqrt{r^2-2rR\cos(\phi-\phi')+z^2}},
\end{equation}
to obtain~\cite{PhysRevD.41.1867, Jaramillo:2010tc}
\begin{equation}
    \label{eq:psi_ellip_integ}
    \Psi = 1 + \frac{M}{2\pi} \left[
    \frac{2\mbox{K}\left( \frac{-4\rho\rho_0}{z^2+(\rho-\rho_0)^2}\right)}{\sqrt{z^2+(\rho-\rho_0)^2}}
    + \frac{2\mbox{K}\left(
\frac{4\rho\rho_0}{z^2+(\rho+\rho_0)^2}\right)}{\sqrt{z^2+(\rho+\rho_0)^2}}
\right],
\end{equation}
where $K(x)$ is the complete elliptical integral of the first kind,
with the convention used by Mathematica 
$$K(x) = \int_0^{\pi/2} \frac{d\theta}{\sqrt{1-k
\sin^2\theta}}.$$
While initial data corresponding to arbitrary ring radii is easy to
construct, actual evolutions of these data are numerically challenging
for large radii, as explained below. Note that $\psi \sim \ln R$,
where $R$ is the coordinate distance to the ring, in the neighborhood
of the ring singularity. Hence, the evolution variable $W =
\psi^{-2}$ will have the form $W\sim 1/(\ln R)^2$, which is
continuous, but not differentiable at $R=0$.

{\it Discrete Ring}:
To construct initial data for the discrete BH ring (i.e. a symmetric
distribution of BHs on a ring) we superimposed $N$ BHs, where the total
mass of the ring is $1M$.
Here $\psi$ is given by
\begin{eqnarray}
    \label{eq:pot_N-BHs}
    \psi &=& 1 + \sum_{i=0}^{N-1} \frac{m_i}{|\vec r - \vec r_i|} \\
         &=& 1 + \sum_{i=0}^{N-1}
\frac{m_i}{\sqrt{(x-x_i)^2+(y-y_i)^2+(z-z_i)^2}},
    \nonumber
\end{eqnarray}
where $m_i = M/N$,  $\vec r_i = [R \cos(i \alpha), R\sin(i \alpha),
0]$ is the coordinate location
of BH $i$, $R$ is the radius of the ring, and $\alpha = 2\pi/N$.
In order to preserve reflection symmetry, $N$ must be even.

For this configuration,
the ADM mass of the spacetime is approximately $M_{\rm ADM}= N m$ and hence the
final merged BH should have a mass of $M\sim N\ m$, where
$N$ is the number of BHs and $m$ is mass of each BH. Binding
energy, which tends to zero as the configuration size goes to
$\infty$,  will reduce the ADM mass by a small amount 
(for example, the binding
energy of a non-spinning BH binary near the ISCO is only about
$2\%$), and gravitational radiation will further reduce the mass of
the final black hole by a small percentage. 

Our technique of using a discrete ring to model the horizon dynamics of 
a continuum ring distribution is a natural extension of the techniques
developed in~\cite{Abrahams:1992ru}.

\begin{figure}
        \includegraphics[width=0.9\columnwidth]{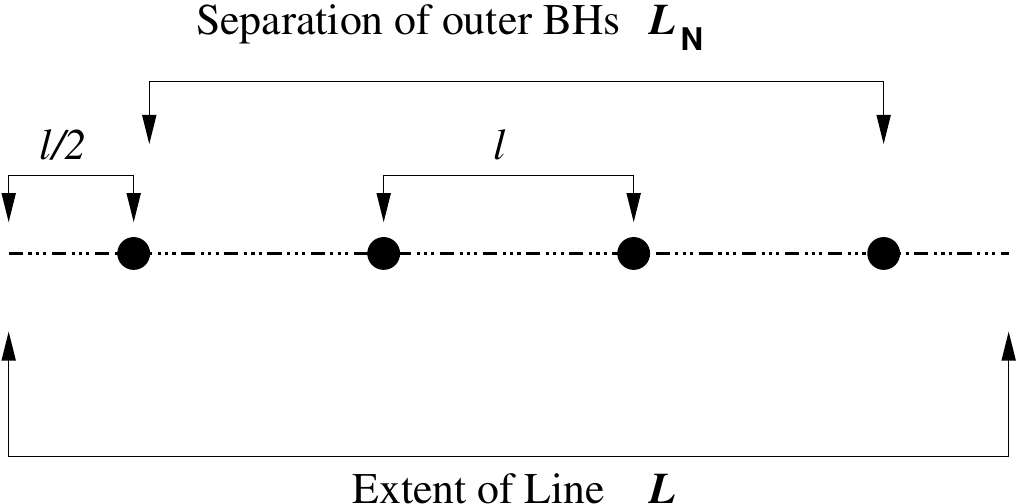}
        \caption{The setup for the linear distribution of BHs.
          $l$ is the separation between neighboring BHs, $L_N$ is the 
          separation between the two outer BHs, while
          $L$ is the length of the line. The line if formally 
          a length $l$ longer than the separation of the two outer
          BHs so that the linear density is the same in the
          neighborhood of each BH (we consider that the
          BH's mass is spread out over an interval of $\pm l/2$ around
          the BHs center.}
        \label{fig:line}
\end{figure}

{\it Line}:
In order to mimic the differential line element of the ring, we consider
a finite line element constrained such that the average linear mass
density of black holes per unit length of the line 
 is the same
as that of the ring.
To construct this linear configuration of BHs, we consider a line of 
length $L$, and place $N$ BHs on the line with a uniform separation
$\ell$. The outermost BHs are arranged a distance $\ell/2$ from the
ends of the line. In this configuration, the mass density of the line
is $m/\ell = M/L$, where $m=M/N$ (see Fig.~\ref{fig:line}).
If $L$ is the total length of the line and $L_N$ is the separation between
 the two outer BHs, then $L=[N/(N-1)] L_N$ and the average linear
density is $M/L = (N-1) m /L_N$.
Hence
\begin{equation}
    \label{eqn:line_length_N}
    L_N = 2 \pi R \frac{N-1}{N},
\end{equation}
where  $R$ is the radius of the ring.

Note this configuration was first studied in
Ref.~\cite{Abrahams:1992ru} to show that a common apparent horizon
does not exist for lines larger than $L\sim1.5M$, and to therefore
argue that this configuration has a naked singularity.

Because ${\cal H}$ is a global entity, the legitimacy of using
 a linear mass distribution to model the behavior of the EH for a ring
is not clear. As we argue below, the EH must be very close to
the linear singularity when the line is sufficiently long, and hence
its structure is, at least partially, dependent on the same singular
behavior in the metric as the EH around the ring singularity. In
addition, the spacetime near the singularity rapidly approaches
flat space in both configurations. So outgoing null
generators far enough away from the singularity to see the differences
in the metric are  also likely to be far enough away to escape to infinity.

The study of the EH structure from a linear distribution
is interesting in its own right, and if we can show that this
distribution does not have a non-singular horizon in the distant past
(which is supported by the work of~\cite{Abrahams:1992ru}),
this helps support the conjecture that any linear distribution, 
sufficiently extended in space, is surrounded by a singular horizon,
or no horizon at all.

\section{Numerical Techniques}
\label{sec:num}
We evolved these BH configuration using the {\sc
LazEv}~\cite{Zlochower:2005bj} implementation of the moving puncture
approach~\cite{Campanelli:2005dd, Baker:2005vv}.
We obtain accurate horizon parameters by
evolving this system in conjunction with a modified 1+log lapse and a
modified Gamma-driver shift
condition~\cite{Alcubierre02a, Campanelli:2005dd}, and an initial lapse
$\alpha(t=0) = 2/(1+\psi_{BL}^{4})$, we also used $\alpha(t=0) =1$ for
the continuum ring.  The lapse and shift are evolved
with
  \begin{eqnarray}
\label{eq:gauge}
(\partial_t - \beta^i \partial_i) \alpha &=& - 2 \alpha K,\\
 \partial_t \beta^a &=& 3/4 (\tilde \Gamma^a - \tilde \Gamma^a(0).
 \label{eq:Bdot}
 \end{eqnarray}
When searching for EHs, we evolved these configuration using the
unigrid PUGH driver~\cite{cactus_web}. After saving the metric at each
timestep, we used the {\sc EHFinder} thorn~\cite{Diener:2003jc} to
locate the event horizons. In this case, we used the PUGH driver
because the publicly available version of {\sc EHFinder} was not
compatible with the {\sc Carpet} AMR driver~\cite{Schnetter-etal-03b}.
When evolving the continuum ring configuration, we used the {\sc Carpet}
diver because the resolutions required near the ring would have made a
unigrid simulation prohibitively expensive.  We used the  {\sc
AHFinderDirect}~\cite{Thornburg2003:AH-finding} thorn to locate
apparent horizons.

In order to make the unigrid runs more efficient, we mimicked
fixed-mesh-refinement using a multi-transition FishEye
transformation~\cite{Baker:2001sf, Alcubierre02a, Zlochower:2005bj}.
In addition to this, we also exploited `octant symmetry' in most of
our simulations, allowing us to
increase the resolution.

Solving the Einstein Equations for a BH ring of arbitrary radius
numerically is not feasible due to resolution constraints. To
understand why this is, we can consider the case of a discrete ring
of some large radius. As we add more BHs to the ring, the mass of each
BH is reduced. The resolution required to evolve a BH is inversely
proportional to its mass, hence in the continuum limit, we would need
an infinitesimal gridspacing. On the other hand, if a spheroidal
common apparent horizon is present, then we only need to resolve the
region outside the AH. Thus, we can evolve ring configurations with
small radii.

When searching for EHs using EHFinder~\cite{Diener:2003jc} we first
perform a standard forward in time evolution, outputting the full 4-d
metric at every timestep, until the remnant BH is
nearly spherical. We then perform a backwards in time evolution and
track the null generators of the EH from the final AH backwards in
time. As noted above, once two generators cross they leave the EH.
Thus we need to remove these generators for all timesteps prior to
their crossing. In practice, we do this by tracking the separation of
each pair of generators, removing the pair if they get within a
predetermined tolerance $\delta$ of each other. In practice we used
$\delta\sim 10^{-4}-10^{-5}$ (depending on the number of punctures).
This tolerance is at least two or three orders of magnitude smaller
than the spatial resolution used for evolving the system.

\section{Results}
  \label{sec:Results}

\subsection{Continuum Ring}
  For the continuum ring we evolved configurations with increasing
ring radii using a set of fixed nested grids (FMR) with the Carpet mesh
refinement driver.

\begin{figure}[!]
        \center
        \includegraphics[width=0.9\columnwidth]{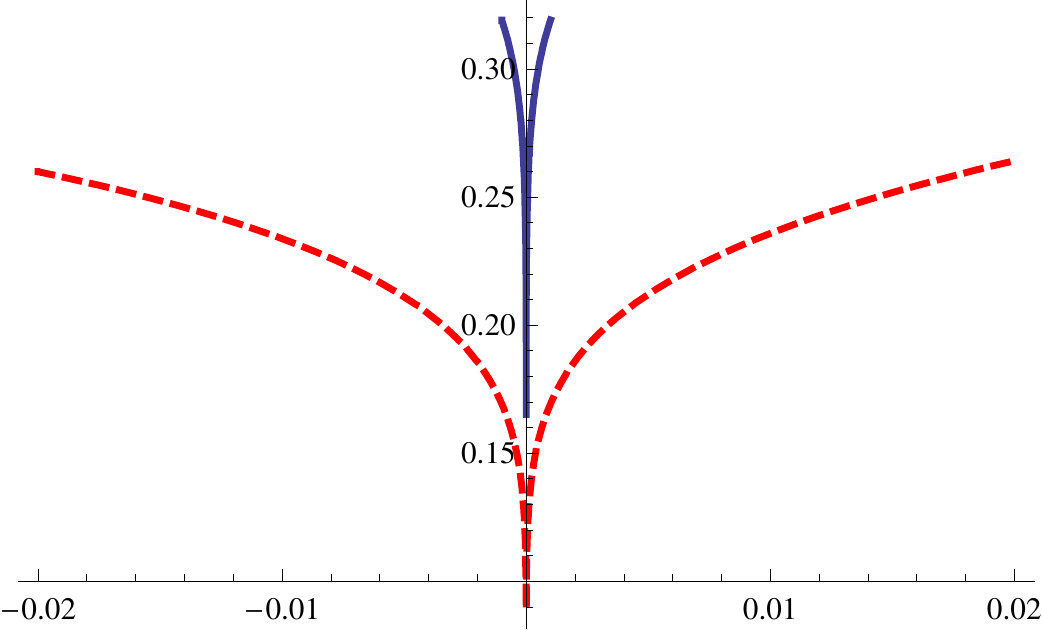}
	\caption{A plot of $W=1/\psi^2$ versus $\tilde \rho=\rho-\rho_0$
         on the $xy$ plane for a ring of radius 1M (dashed) and
  2M (solid). Note how much smaller the region $W<0.3$ is for the
larger ring radius. Based on the figure we estimate that an order 100
  times the resolution is required to properly resolve the $r=2M$
ring.
        }
        \label{fig:psi_comp}
    \end{figure}

  We found that the resolution required to find the AH on the initial slice
increases with the size of the ring (keeping the mass fixed). We were
able to find AHs on the initial timeslice for rings with radii as
large as $0.5M$. 
Based on axisymmetric simulations~\cite{PhysRevD.41.1867, Jaramillo:2010tc},
 we know that AHs exist
for rings of larger radii, but we were not able to find
these initial AHs in our full 3d simulations due to a lack of
spatial resolution. Even though we were unable to find an initial
AHs numerically, we were still able to evolve ring configurations
with radii as large as $~1M$. On the other hand, an
evolution of a ring with radius larger than $r\approx 1.2 M$
was not possible even at high resolutions. Essentially,
the logarithmic singularity in the metric disappears 
numerically due to the effects of finite differencing. That is,
the evolution variable $W\sim 1/\log^2 R$, which is not
differentiable. As is, this would not be a significant problem, but
for large rings, the volume where $W$ is close to zero is very small
(see Fig.~\ref{fig:psi_comp}).
Consequently $W$ gets smoothed out in this region by finite-difference
errors, and the central object quickly loses mass and disappears (and
subsequent evolution show no evidence of collapse or an apparent
horizon). For intermediate radii of $0.5<r\leq 1.0M$, we were able to
evolve the ring. In these
cases, the ring collapsed and eventually a common AH was found. 

    \begin{figure}[!]
        \center
        \includegraphics[width=0.9\columnwidth]{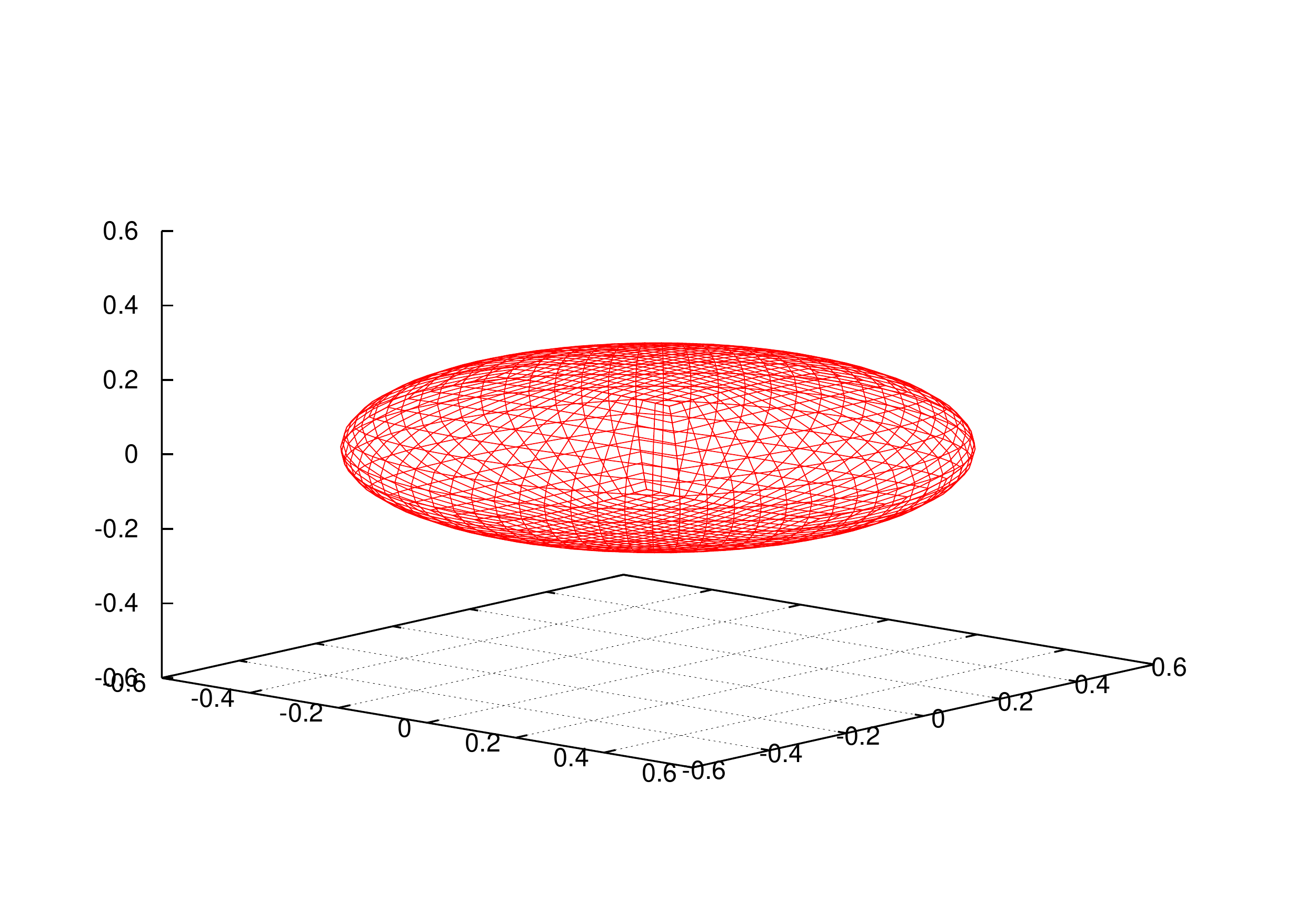}
	\caption{An early apparent horizons for a
``continuum ring'' singularity of radius 0.5 and unit mass.
  The AH is clearly an oblate spheroid with largest radius on
  the $xy$ axis, corresponding to the plane of the ring singularity.
        }
        \label{fig:AHs_cont-ring_radius-0.5}
    \end{figure}

  We used a 3d-Cartesian coordinate grid with outer boundaries at
$100M$, with 3 ghost and 3 buffer zones, 8 levels of refinements.
  We found that the apparent horizons for the ring singularity are
oblate spheroids with minor axes through the perpendicular
direction to the plane of the ring, i.e. coincident with the axis of
the ring (Fig.~\ref{fig:AHs_cont-ring_radius-0.5}). We found that for
ring radii larger than $0.5M$, the ring failed to collapse and
essentially evaporated numerically. By changing the initial
lapse to $\alpha(t=0)=1$ we were able to evolve rings with radii as
large as $\sim 1.0M$. However, this changed the AH at
intermediate times from oblate to prolate. This would seem to indicate
that $\alpha(t=0)=1$ leads to a distorted $z$ coordinate during
intermediate times. In all cases , the AH relaxed to a
sphere at late times. We note that the AH area always increased from
its initial value (or the value found at the first time the AH was detected).
We did not search for EH for the continuum ring because we were unable
to evolve rings of sufficient radius to have interesting EHs.

\subsection{Discrete Ring}

We evolved the discrete ring case in order to gain insight into the
behavior of continuum rings with large radii (where numerical
simulations are not feasible). The moving puncture approach has
already been shown to work for large numbers of discrete
punctures~\cite{Campanelli:2007ea, Lousto:2007rj, Galaviz:2010mx}, and
is therefore well suited for simulating the discrete case. The goal
here was to see if, when we increase the number of BHs in the ring,
while keeping the ring mass and radius fixed, a common horizon forms.
In practice, we evolved configurations with $N=2,4,6,8,10,20$ BHs.

For the cases of  $N=10$ and $20$ BHs, the overhead of using a unigrid
setup was too large. We performed some preliminary experiments
calculating AHs using Carpet and FMR. However, we were not able to
search for EHs.  For these cases, we simulated rings with radii as
large as $2.5M$.  As expected, initially there were $N$ distinct AHs,
which then merged into a common AH,

For $N=2,4,6,8$, in order to find the EHs we used a unigrid
setup (a requirement for the EHFinder thorn) with outer boundaries
at a coordinate distance of $12.5M$, which corresponds to a physical
distance of $45M$ due to the FishEye coordinates.
Because the BH mass, and hence the required maximum gridspacing,
scales as $1/N$, we used resolutions of $h=M/12$ and $h=M/16$ for $N=2$,
$h=M/24$ and $h=M/32$ for $N=4$,
and $h=M/48$ for $N=6,8$.
 In all the
cases the Courant factor was set to $0.5$.
  We varied the radius of the ring from $1.0M$ to $2.5M$.

\begin{figure}[!]
     \includegraphics[height=0.5\columnwidth,angle=-90]{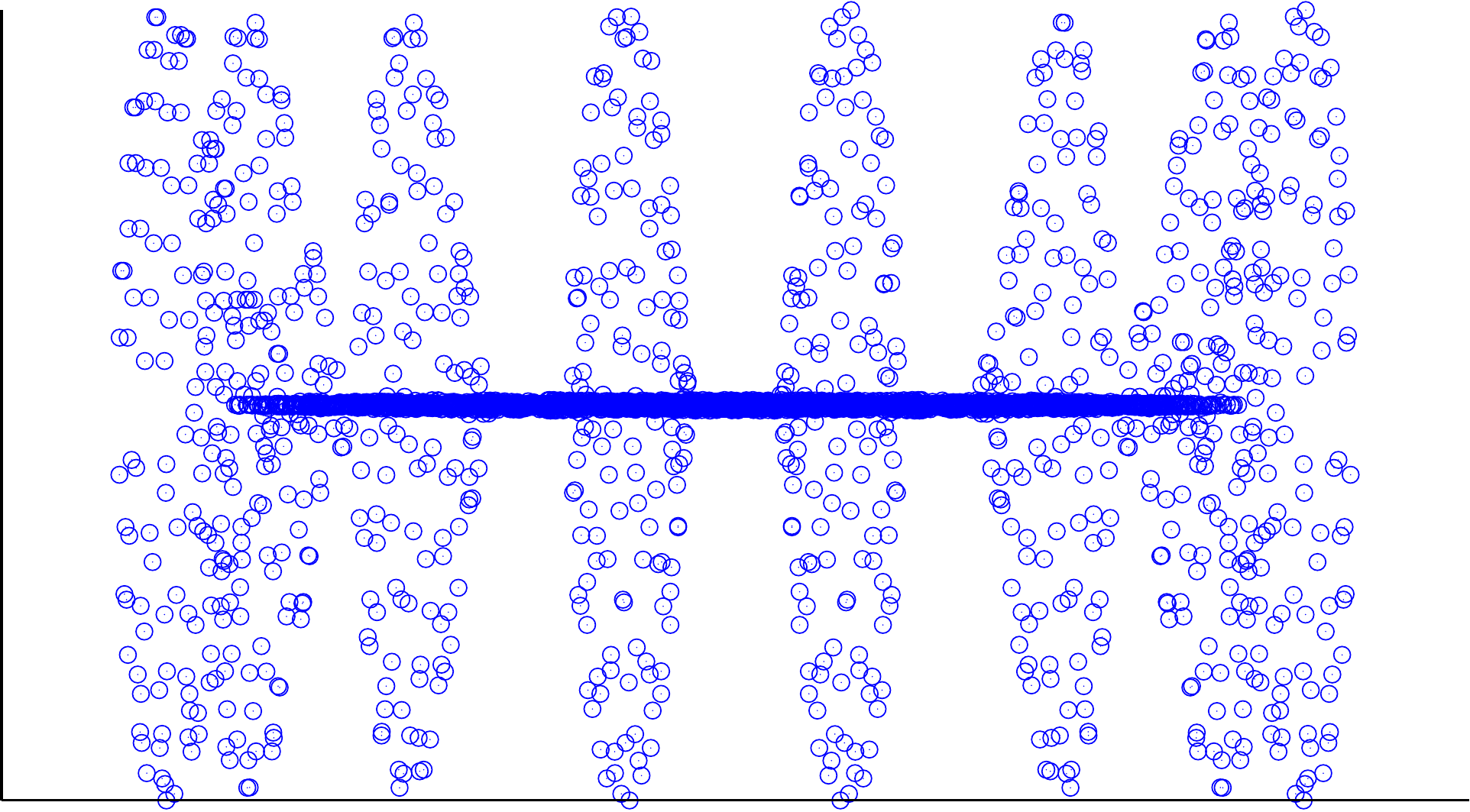}
     \includegraphics[height=0.5\columnwidth,angle=-90]{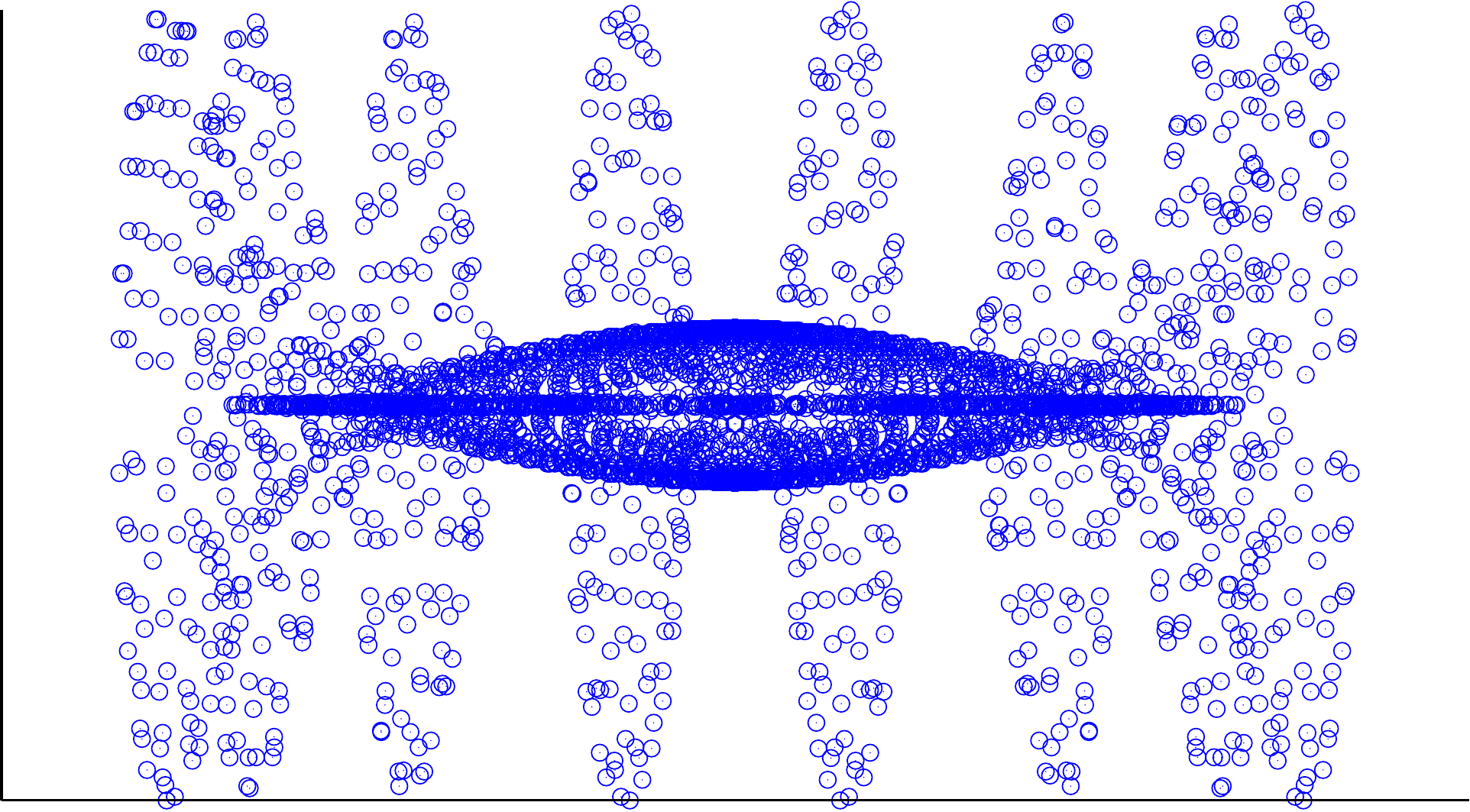}
    \caption{An edge-on view of the `pancaking' of the central part
    of the EH for 8 BHs on a ring. The
figure on the left shows the generators of the horizon at the timestep
when the central generators just pass through the caustic and enter
the horizons.  The figure on
the right shows the generators one timestep later. In the figure on the
left, the central generators are just crossing (thus are not part of
${\cal H}$, while in the figure on the right an extended central
object is visible. Note that the $z$ axis is magnified by a factor of
10 compared to the $x$ and $y$ axes.}
    \label{fig:8BH_EH_pancake}
  \end{figure}

\begin{figure}[!]
     \includegraphics[width=0.33\columnwidth]{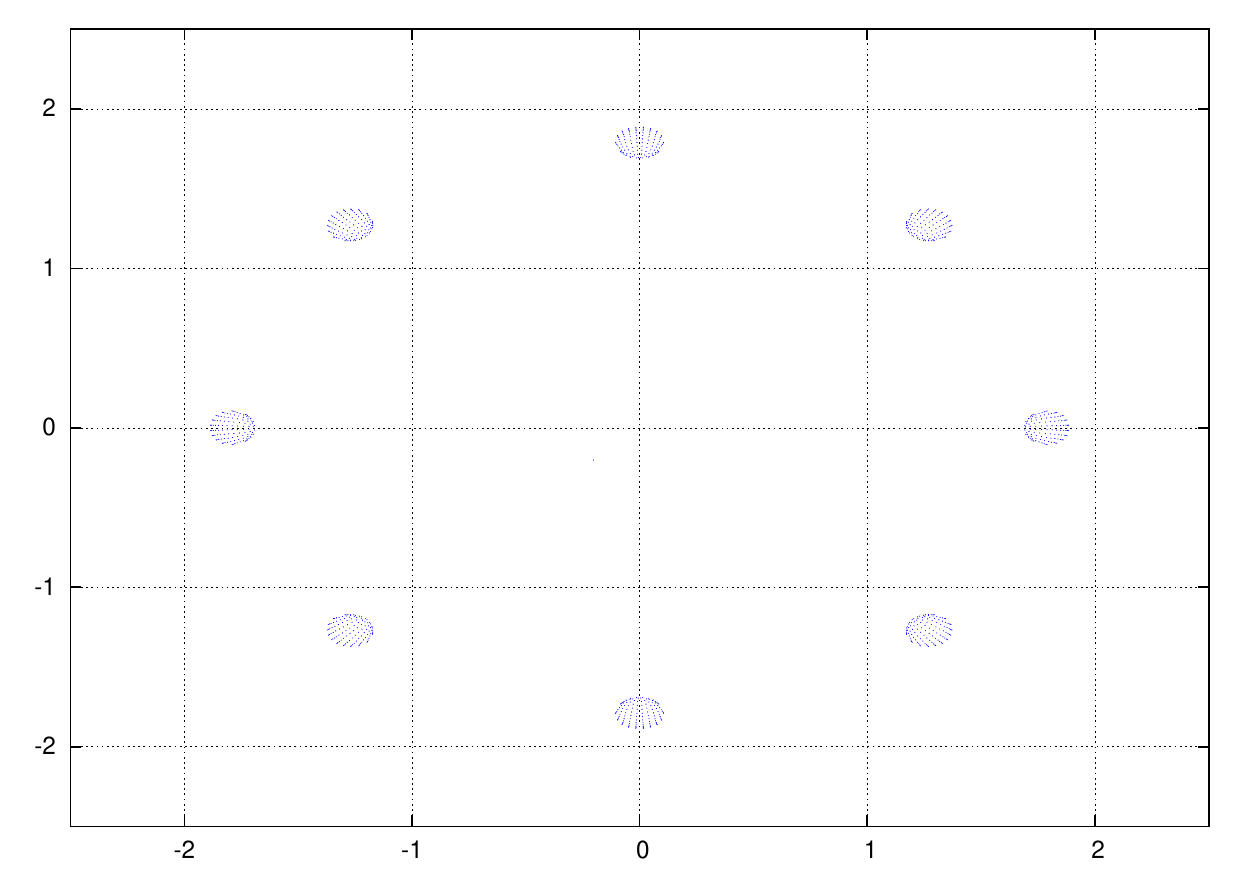}
     \includegraphics[width=0.33\columnwidth]{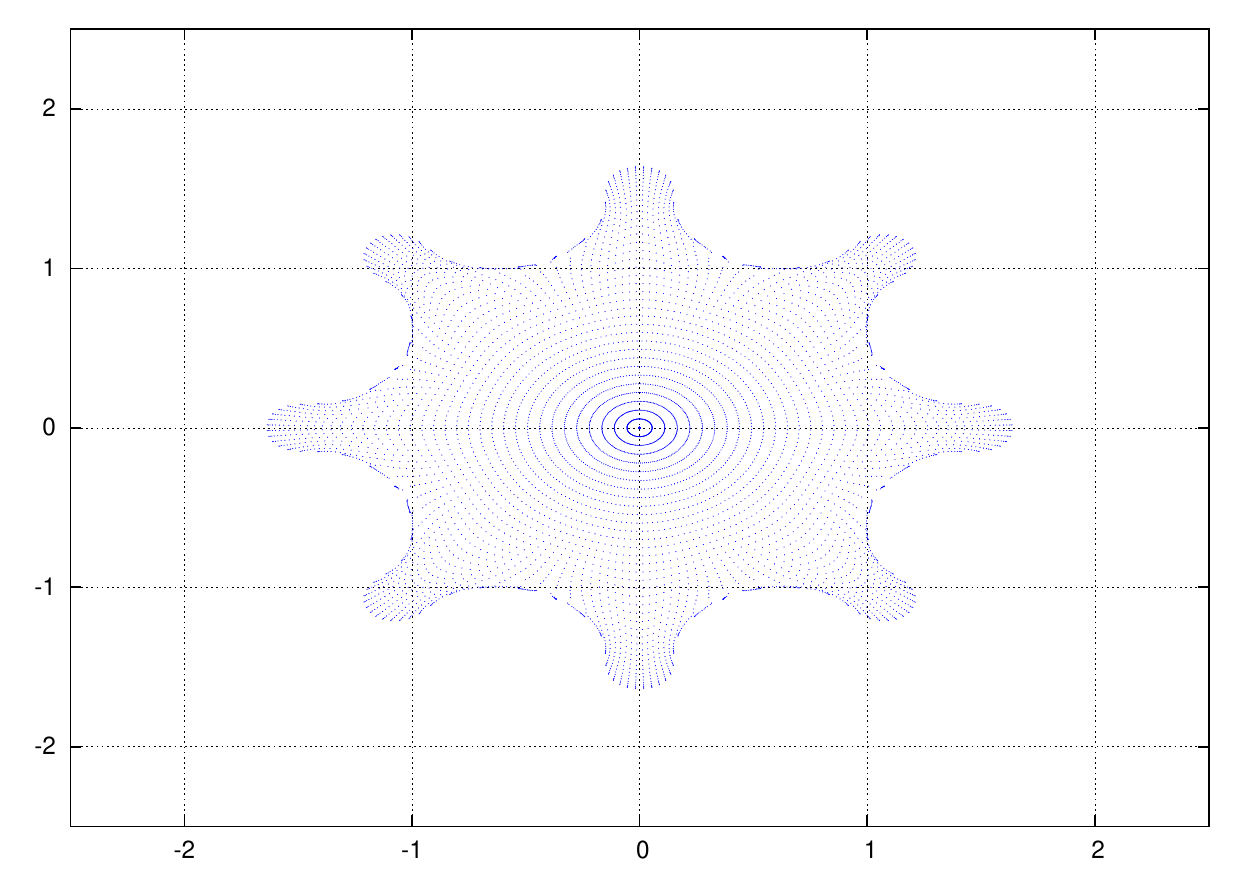}
     \includegraphics[width=0.33\columnwidth]{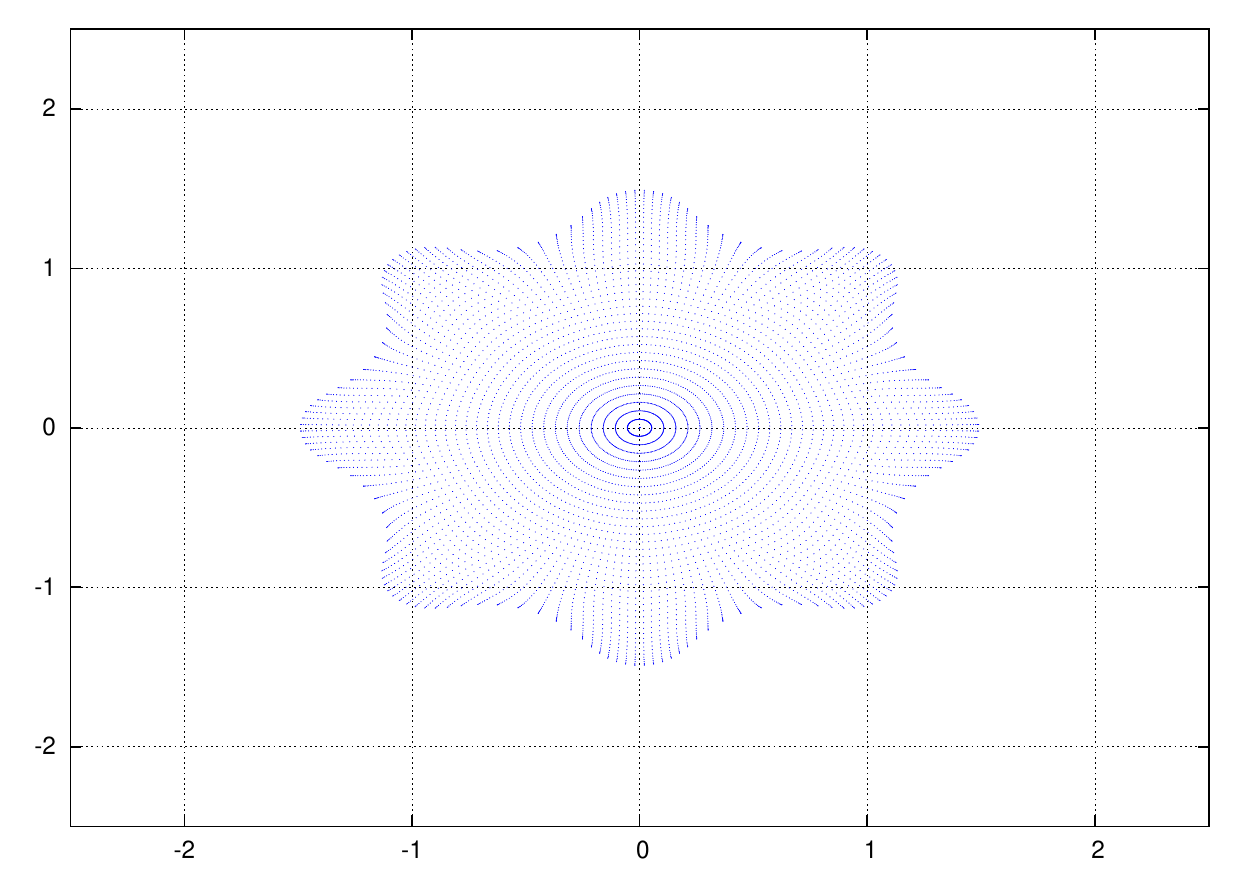}
    \caption{$xy$ projections of the 8 BH ring configuration showing
a time sequence with (left) 8 individual EHs, (center) a highly distorted
common EH, (right) and a less distorted common EH.}
  \label{fig:8BH_xy_proj}
\end{figure}

\begin{figure}[!]
     \includegraphics[width=0.33\columnwidth]{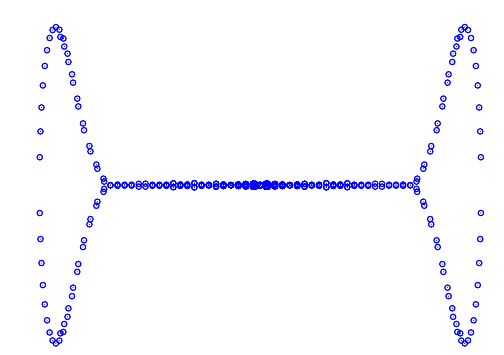}
     \includegraphics[width=0.33\columnwidth]{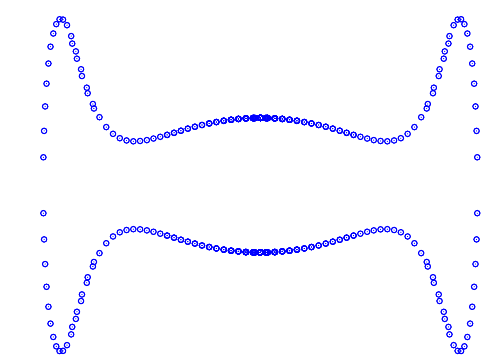}
     \includegraphics[width=0.33\columnwidth]{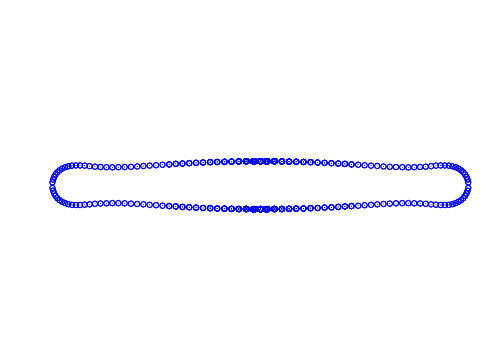}

     \caption{$xz$ cuts of EH for the 8BH ring configuration
  at the instant in time when the
common EH forms (left), two timesteps later (center, note the concave
shape of the central region), and five timesteps after that (right).
Note that the $z$ axis in the two leftmost figures is magnified by
a factor of 10 compared to the $x$ axis, 
while the $z$ scale is equal to the $x$ scale in the last
figure.}
    \label{fig:8BHS-EHs_cuts_XZ}
  \end{figure}

\begin{figure}[!]
     \includegraphics[height=0.50\columnwidth,angle=-90]{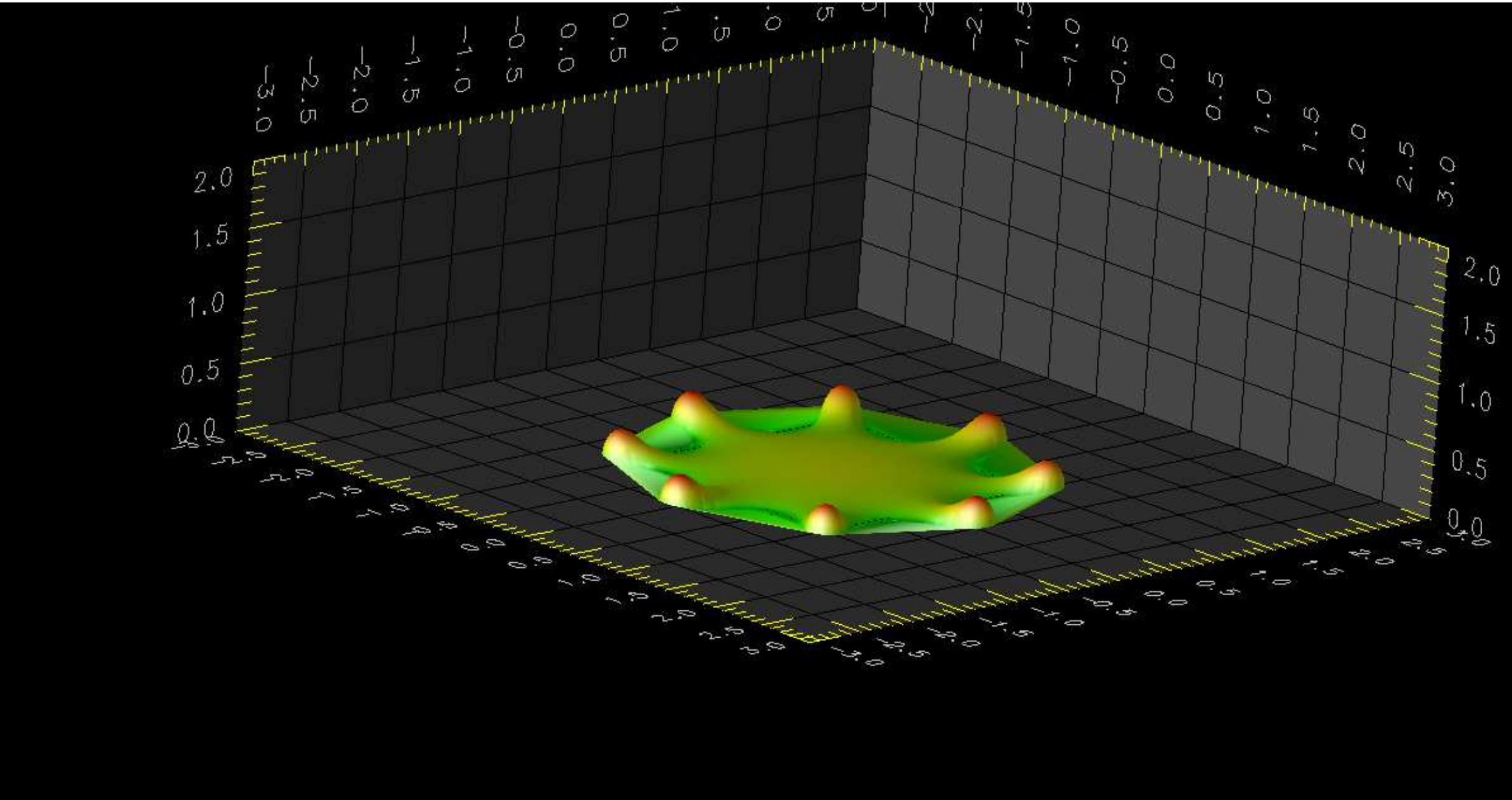}
     \includegraphics[height=0.50\columnwidth,angle=-90]{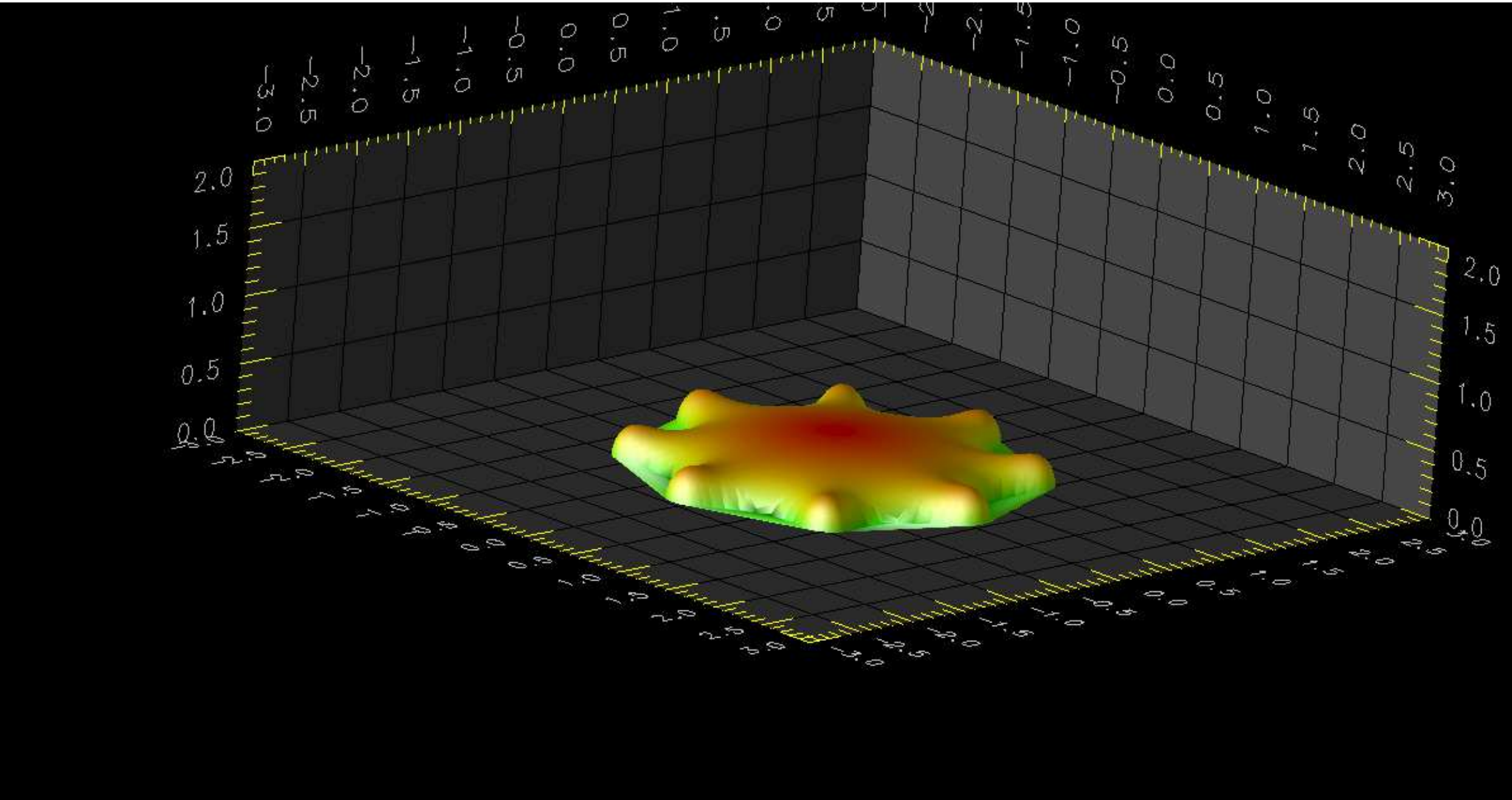}

     \includegraphics[height=0.50\columnwidth,angle=-90]{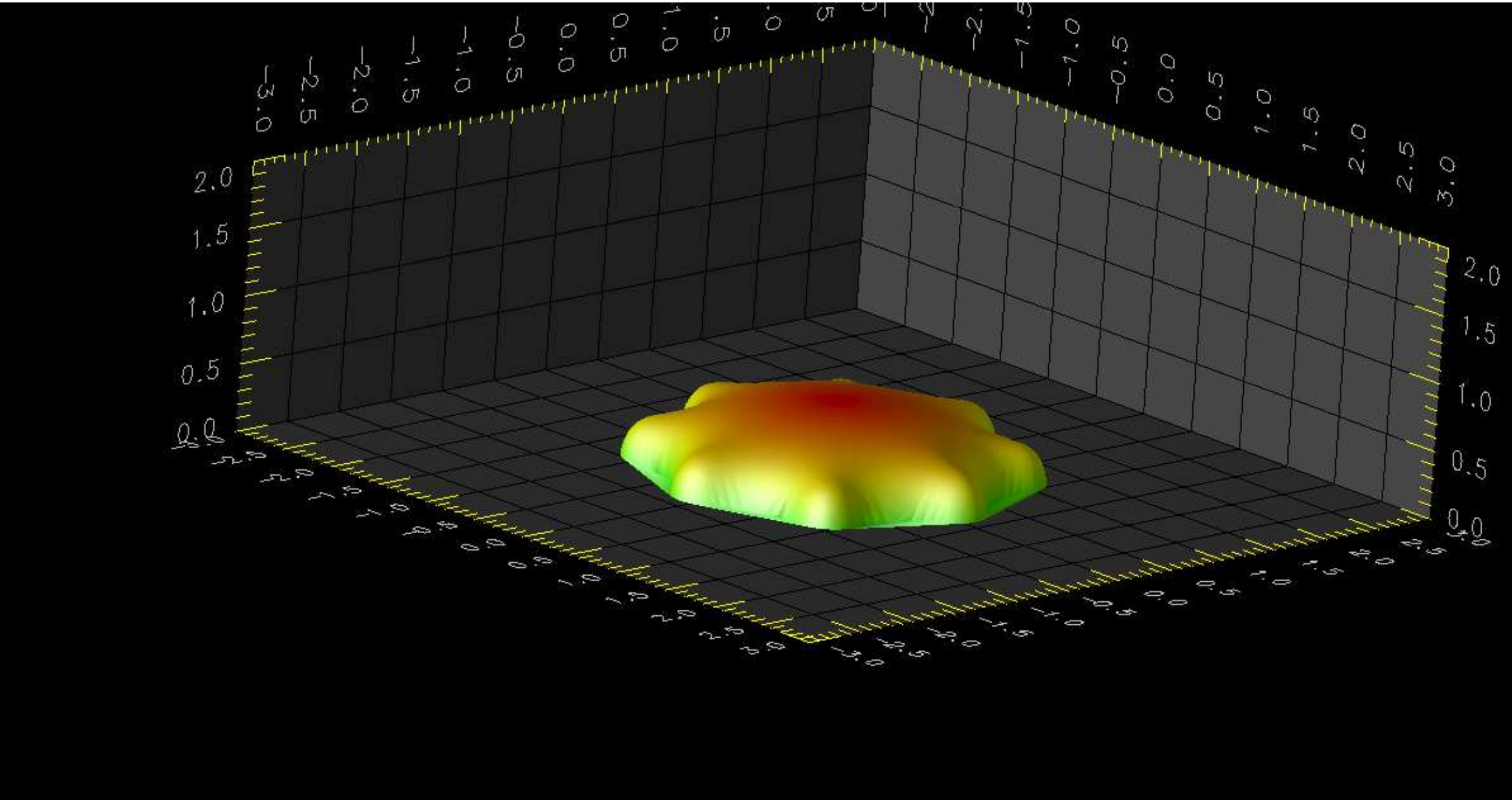}
     \includegraphics[height=0.50\columnwidth,angle=-90]{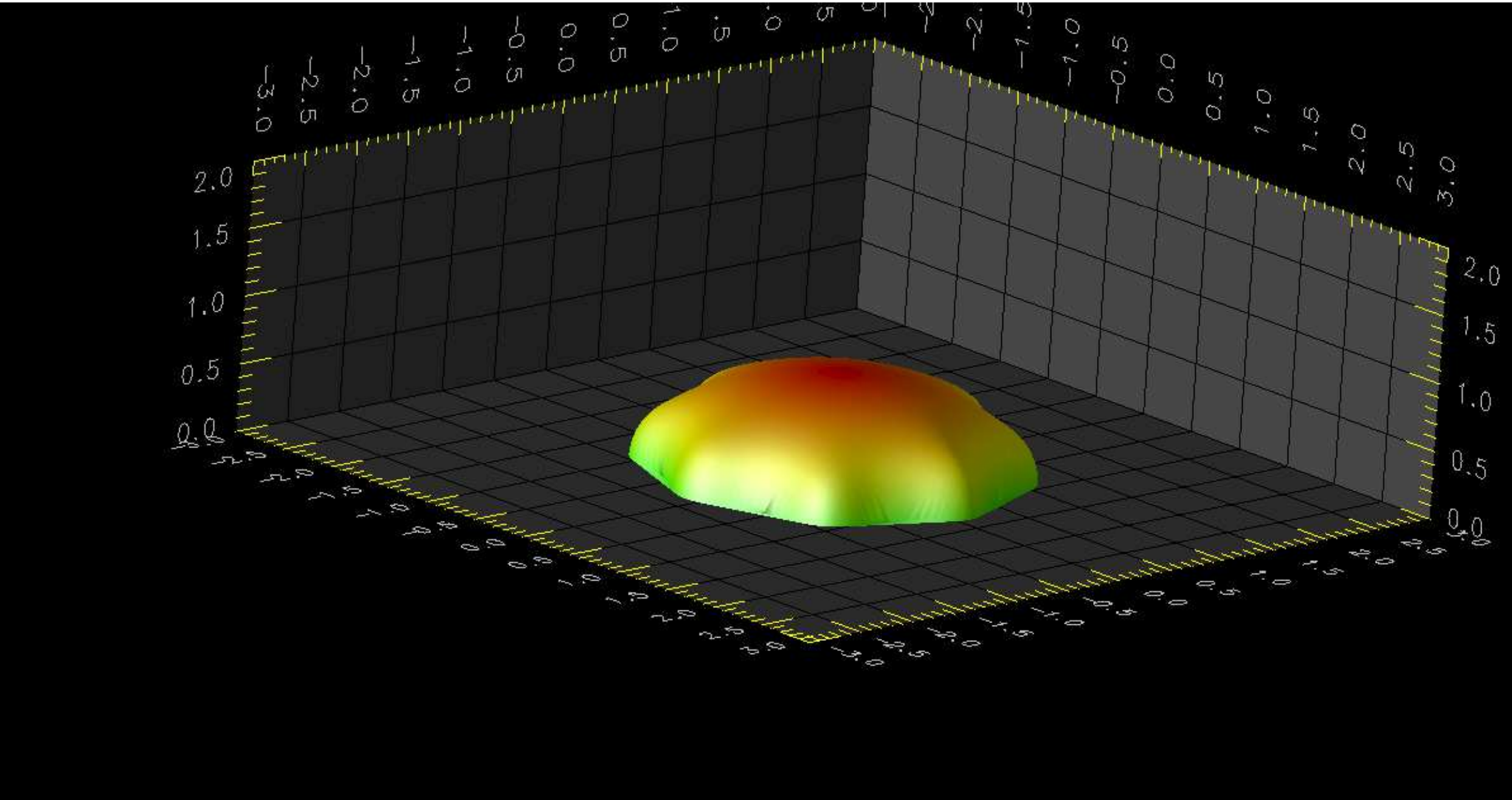}

     \includegraphics[height=0.50\columnwidth,angle=-90]{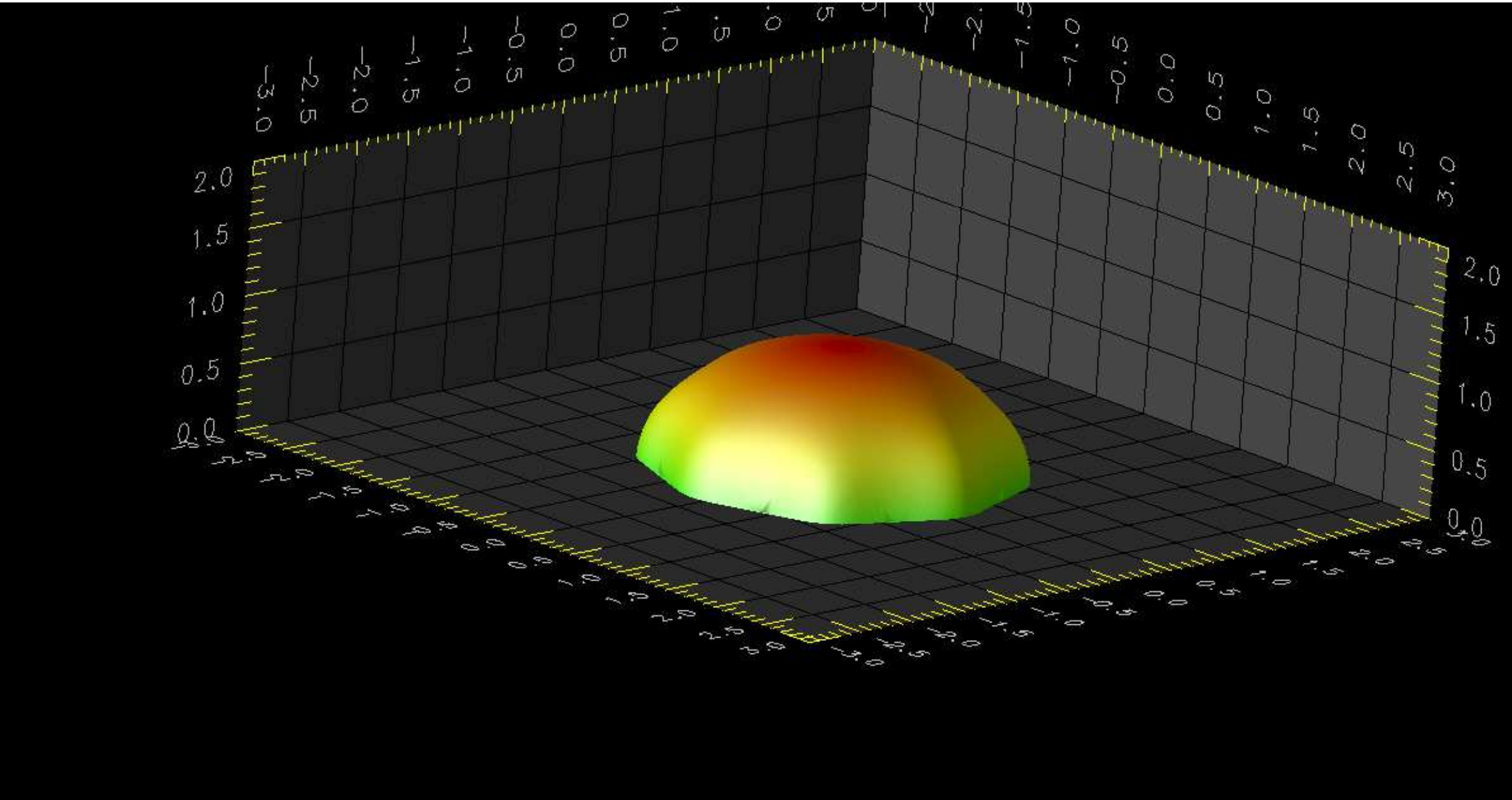}
     \includegraphics[height=0.50\columnwidth,angle=-90]{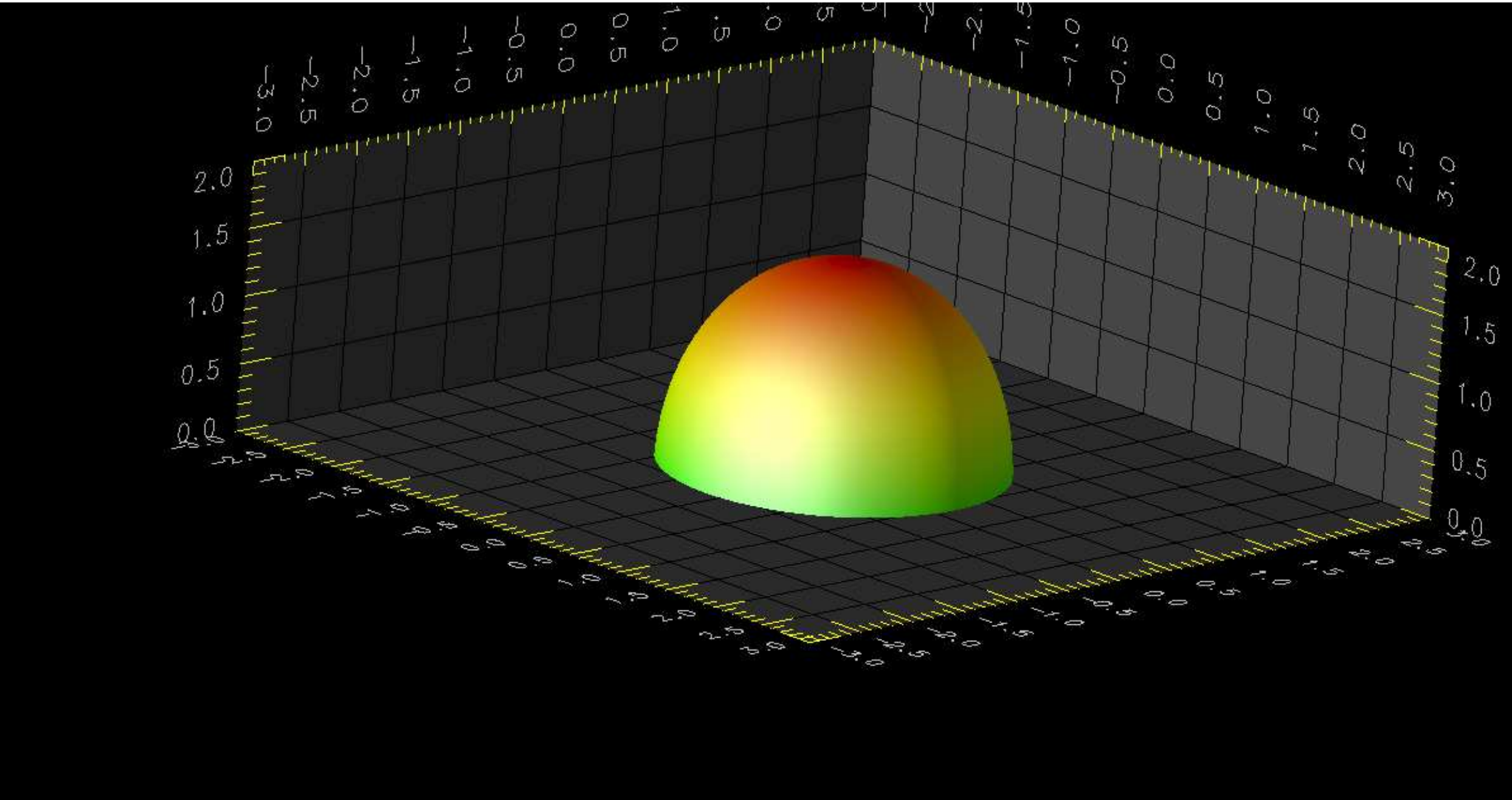}
     \caption{3-d snapshots of the 8 BH ring configuration (only the top
      part of the horizon ($z\geq0$ is shown).
      The sequence in time runs from left to right and top to bottom,
      beginning with the first common horizon and forward in time from 
      then to when the horizon in nearly spherical. Note that the flat
      (green) sheet apparent between the horizon on the first and
      second slices is an artifact of the visualization and does not
      belong to the horizon.
      This provides another way to visualized the `pancaking' process
      also visible in
      Figs.~\ref{fig:8BH_EH_pancake}~and~\ref{fig:8BHS-EHs_cuts_XZ}.}
    \label{fig:ringconfig_8BHS-3d}
  \end{figure}

\begin{figure}[!]
\begin{center}

     \includegraphics[height=0.75\columnwidth,angle=-90]{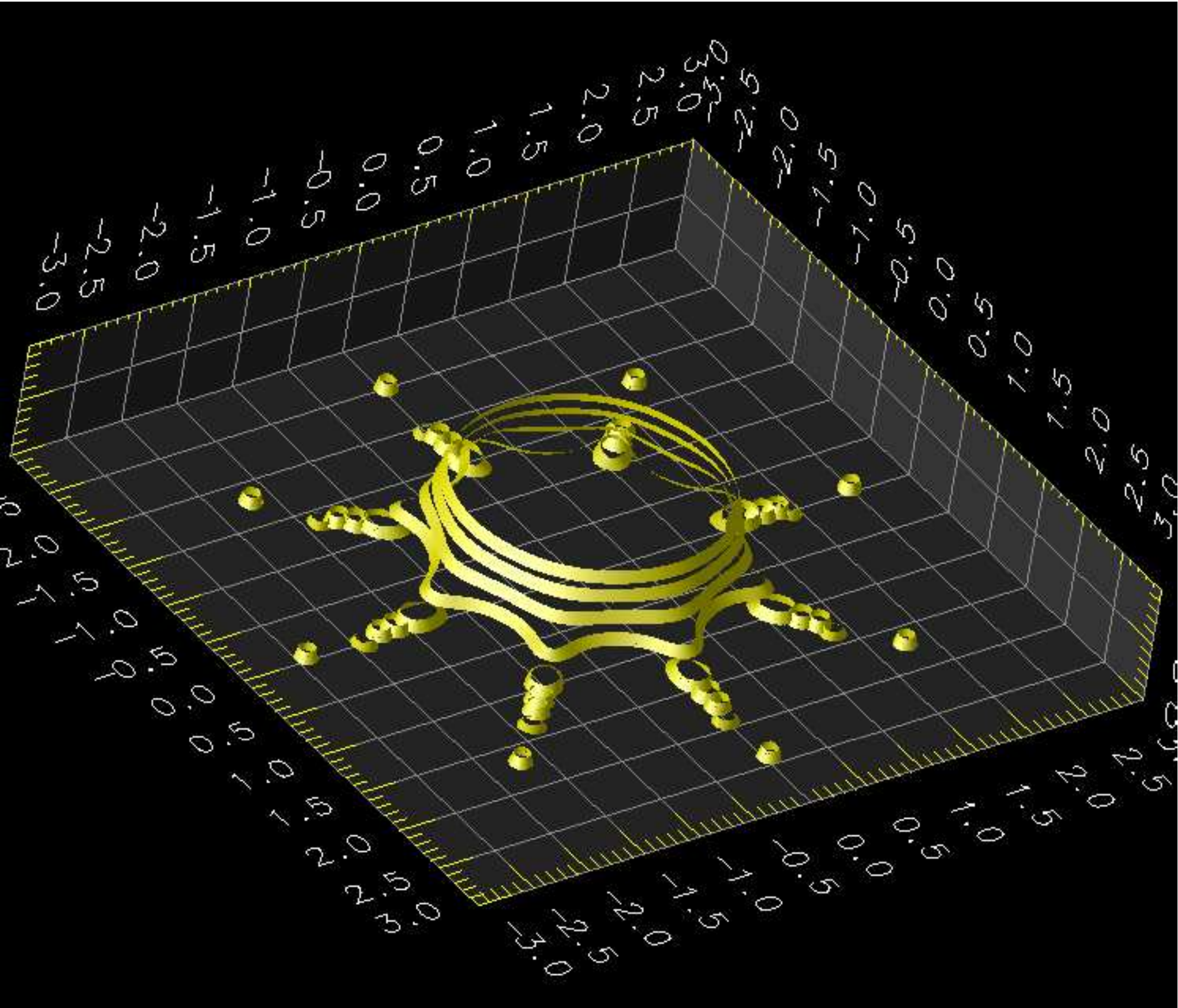}
\end{center}
     \caption{ A sequence of $t=const$ slices of EH for the 8 BH ring
configuration arranged vertically to show the ``octopus'' like structure 
of the EH}
    \label{fig:ringconfig_8BHS-octopus}
  \end{figure}

In Fig.~\ref{fig:8BH_EH_pancake} we show the 8-BH discrete ring EH at
the instant when the horizon transforms from  8 distinct objects to a
single distorted horizon of topology $S^2$, as well
as the horizon at one timestep later when there is only a single EH.
As noted above, the EH is actually found by a backward in time
evolution. From this perspective, we note how the central part of the
horizon `pancakes' to zero width during a single timestep.
Also note that the central part of the horizon is concave, indicating
that the generators near $(x,y) =0$ will not cross first. Thus a
``hole'' will not form and the EH (in this slicing) will not have a
toroidal topology. In Figs.~\ref{fig:8BHS-EHs_cuts_XZ}~and~\ref{fig:8BH_xy_proj}
we show $xy$ projections and $xz$ cuts of the same 8 BH
configuration, while in Fig~.\ref{fig:ringconfig_8BHS-3d} we show
3-d plots of the EHs (for $z\geq0$).
In Fig.~\ref{fig:ringconfig_8BHS-octopus} we show a sequence of
$t=const$ slices of the EH, arranged vertically, to show the
``octopus'' like structure of this EH.
 Similar results for $4$ and $6$ BHs show that no
central ``hole'' forms at any time.


\begin{figure}[!]
    \begin{center}

     \includegraphics[width=0.75\columnwidth]{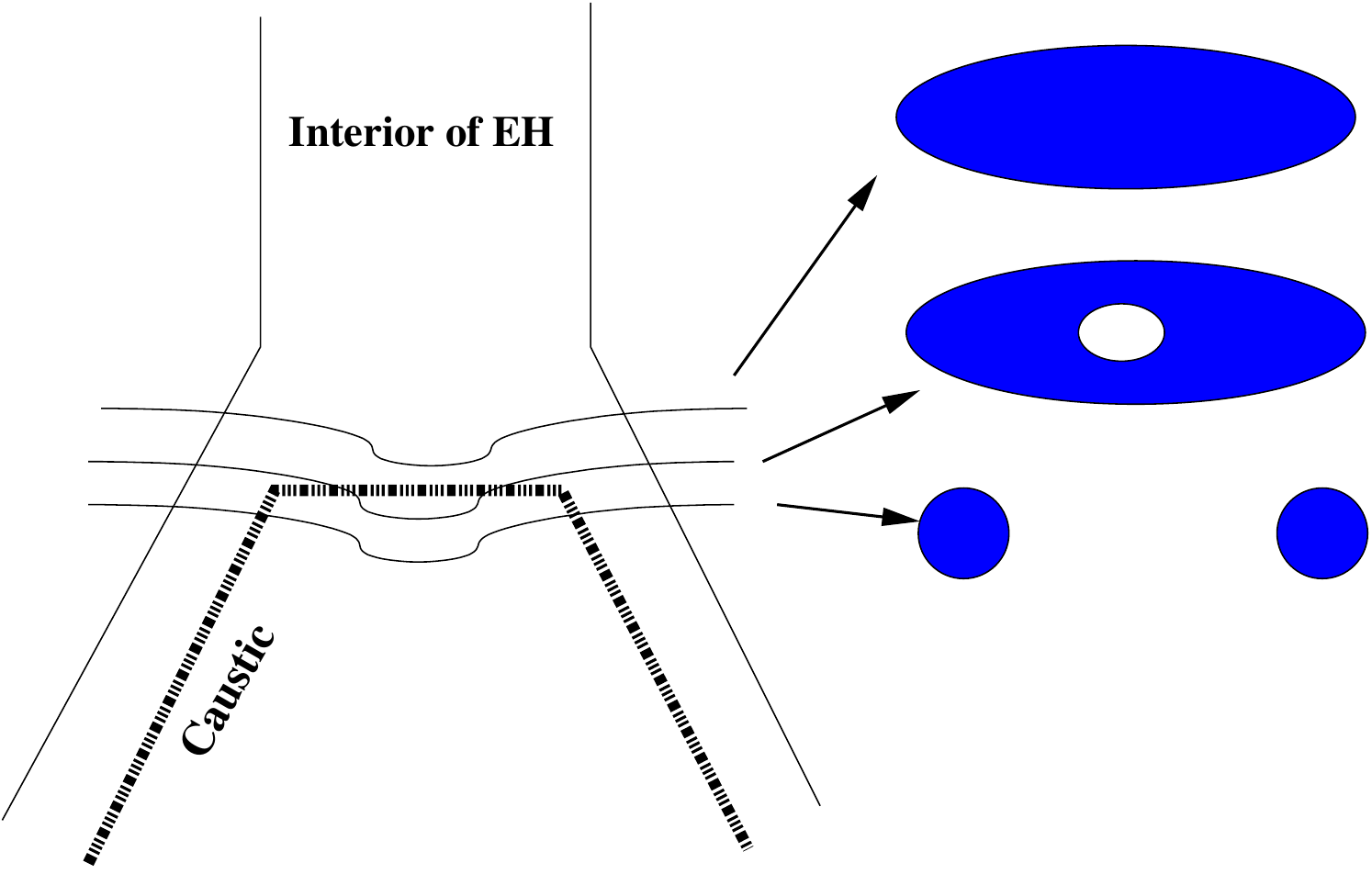}
     \end{center}
     \caption{The caustic structure for the multiple black holes
arranged on a ring. Only the $tx$ plane is shown.
From the figure, one can see that a toroidal horizon is possible if
the slicing near the origin is retarded.}
    \label{fig:Caustic}
  \end{figure}

Although we found no toroidal slice here, the caustic structure of the
horizons indicates that a toroidal slice is possible. That is,
the caustic forms a 2-d spacelike {\it plane}, and a minor distortion
of the slicing should produce a new slice that is slightly more
advanced in a small volume outside the origin than it is at the origin
itself. For example, one could try to modify the right-hand-side of
Eq.~(\ref{eq:gauge}) by adding a term of the form $-f(t) \exp[-(r/\sigma)^2]$,
which should retard the slicing around the origin,leading to a
toroidal slice (see Fig.~\ref{fig:Caustic}).

\subsection{Linear BH}

  As a test of conjecture that a black hole ring is not surrounded by
a finite-sized toroidal horizon (for sufficiently large ring radii),
 we simulated
the case where a differential line element of the ring can be
approximated by a finite line segment. In other words, in order for
there to be a common EH in the discrete ring case (with sufficiently
large ring radius) a single EH must surround two neighboring
punctures as the number of punctures becomes very large. In which
case, a small section of the ring will look like a linear distribution
of BHs. The question then is, if we have a finite length linear
distribution of BHs of fixed total mass, will a common horizon form if
the number of punctures in the line is increased arbitrarily, while
keeping the total mass of the line fixed.
In order to keep the linear mass
density constant as we change the number of BHs on the line ($N$), the
length of the line element is given by
Eq.~(\ref{eqn:line_length_N}).

There are two important objections to our modeling of the ring
distribution with a linear distribution. First,
because the metric on the initial slice is obtained by solving an
elliptical equation, the notion that a linear mass distribution will
mimic the relevant regions of the spacetime near a ring distribution
requires some justification. Second, because the event horizon is a
global structure, the global differences between the two spacetimes
can cause the two event horizons to have very different property.

Our argument for using this linear distribution model goes as follows,
as we move backwards in time and the ring gets larger, the event
horizon must get progressively closer to the ring. This is due to the
fact that the conformal factor $\psi$ approaches $1$ (and hence the
metric rapidly approached Minkowski) progressively more rapidly as the
ring radius is increased (see Fig.~\ref{fig:psi_comp}).  Therefore ${\cal
H}$ gets progressively closer to the ring as the ring radius is
increased. Indeed, because the region where $\psi$ differs
significantly from $1$ shrinks very rapidly as the ring radius
increases, the horizon radius must tend to zero faster than the
reciprocal of the ring radius.
Otherwise the outgoing null generators would enter a region where the
space was nearly Minkowski and, due to the rapid falloff of
$\psi$  would therefore be able to escape to infinity.  Consequently,
for a large enough ring, the horizon will lie in a region where the
metric is dominated by the singular $\log R$ term in $\psi$ and will
look like the metric in the vicinity of a linear mass distribution. As
noted above, due to the global nature of the event horizon, it is not
clear how precise the correspondence is between the horizon structure
of the line and the ring. Consequently our results for the linear
distribution should only be considered suggestive of the behavior of
the horizon around the ring.

With the aim of numerically testing the conjecture that no finite
sized EH exists for the distant past
in  the continuum ring case, we
ran simulations with different values of $N$.  First we determined the
minimal distance between two black holes to generate two isolated
(non-connected) event horizons in their initial configuration. We
found that for a system of 2 black holes with total mass $M=1.0$, the
minimal separation between the BHs is $\approx0.9M$.  Given this, we
then proceeded to evolve configurations for $N=4$ with a line length
of $4.0M$ (which guarantees that no common EH exists initially). We
then increased the number of BHs on the line and measured the
coordinate separation of the two central holes as a function of $N$.
For this analysis, we rescaled this inter-BH separation by the
reciprocal of the individual BH masses.  That is, the separation will
decrease as $N$ increases naturally because there are more BHs on the
line, but we are interested in whether or not the EHs will merge as
$N$ increases. If the EHs merge then the separation $l$ of the EHs
will approach $2 r$, where $r$ is the EH radius. If on the other hand
$l/r$ tends to a constant larger than $2$, as was the case here, then
no common EH will form. Since the BH masses, and therefore coordinate
radii, are proportional to $1/N$, we rescale the distances by a factor
of $N$.

\begin{figure}
  \begin{center}
  \includegraphics[width=0.85\textwidth]{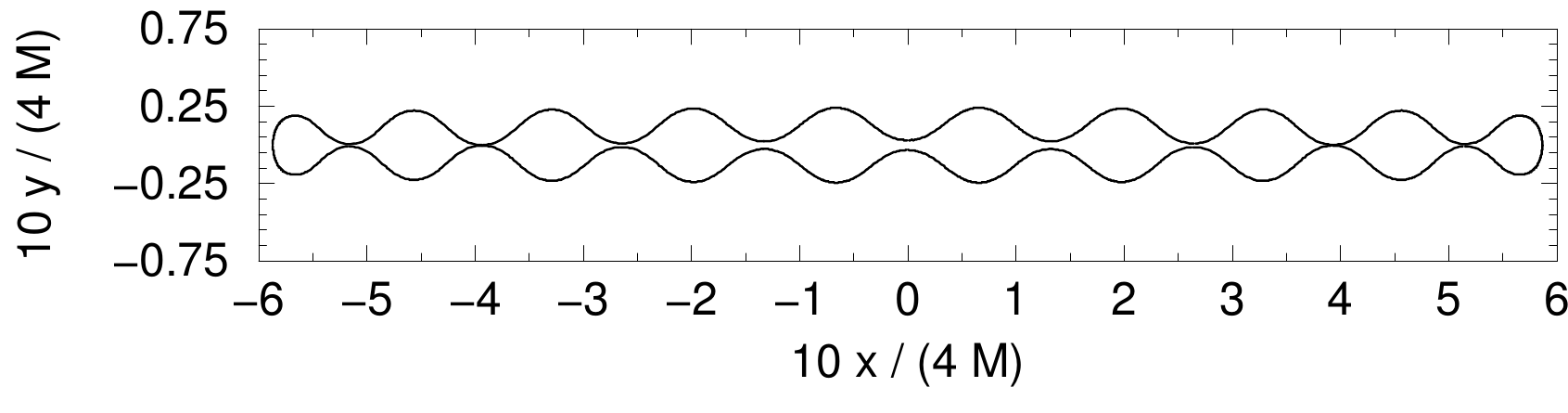}

  \includegraphics[width=0.85\textwidth]{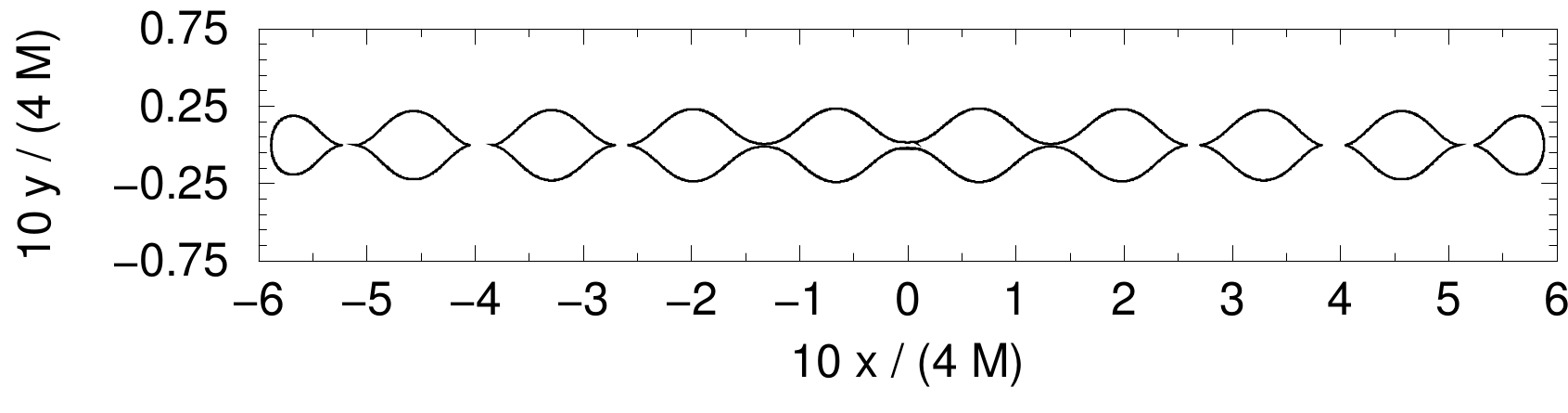}

  \includegraphics[width=0.85\textwidth]{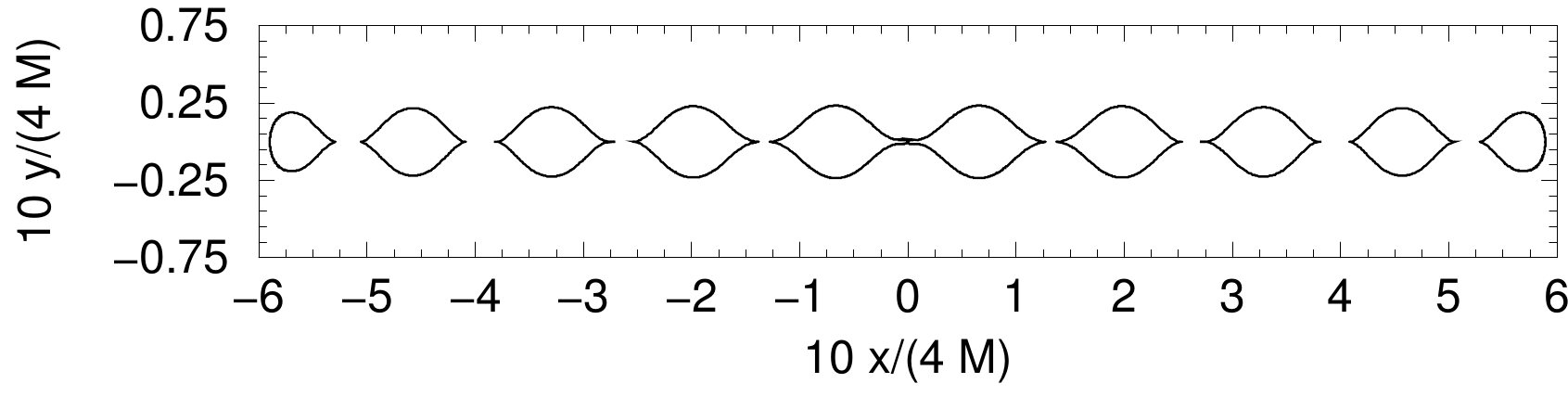}
  \end{center}
  \caption{$t={\rm const}$ slices of the
     intersection of ${\cal H}$ and the $xy$ plane for the 10 black
hole line configuration. The figure on top
    shows the first common EH, the figure in the center
shows an earlier slice with 7 distinct objects, while the figure on
the bottom shows the EH when there are 9 distinct objects. Note that
the two central BHs merge first.}
  \label{fig:line_10BHs}
\end{figure}

In Fig.~\ref{fig:line_10BHs}, we plot the intersection of ${\cal H}_t$
with the $xy$ plane for the 10 BH line configuration for several
different times. Interestingly, the two central BH merge first,
followed by the two BHs neighboring the central ones, followed by the
outermost ones. At one time there are three distinct EH formed by the
two outermost EH on either side (2 objects) of the line and the 6 central
objects (1 object).

\begin{figure}
  \center \includegraphics[width=0.85\columnwidth]{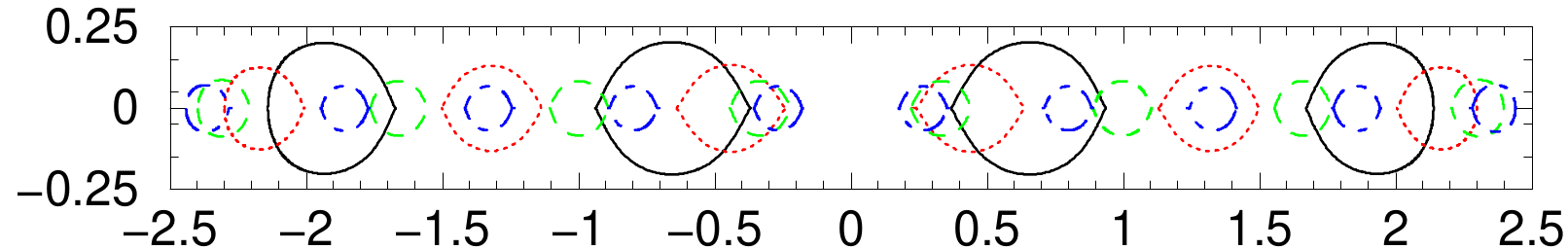}
  \caption{Sets of EHs near $t=0$ for the linear
distribution with $N=4$ (solid), $N=6$ (dotted), $N=8$ (dashed),
  and $N=10$ (dot-dashed) BHs. In this plot the coordinates have not
been rescaled. Note that as $N$ increases the two innermost horizons
approach each other, but also simultaneously shrink in size.
}
  \label{fig:line_allBHs-noscaled}
\end{figure}

\begin{figure}
  \begin{center}
  \includegraphics[width=.45\textwidth]{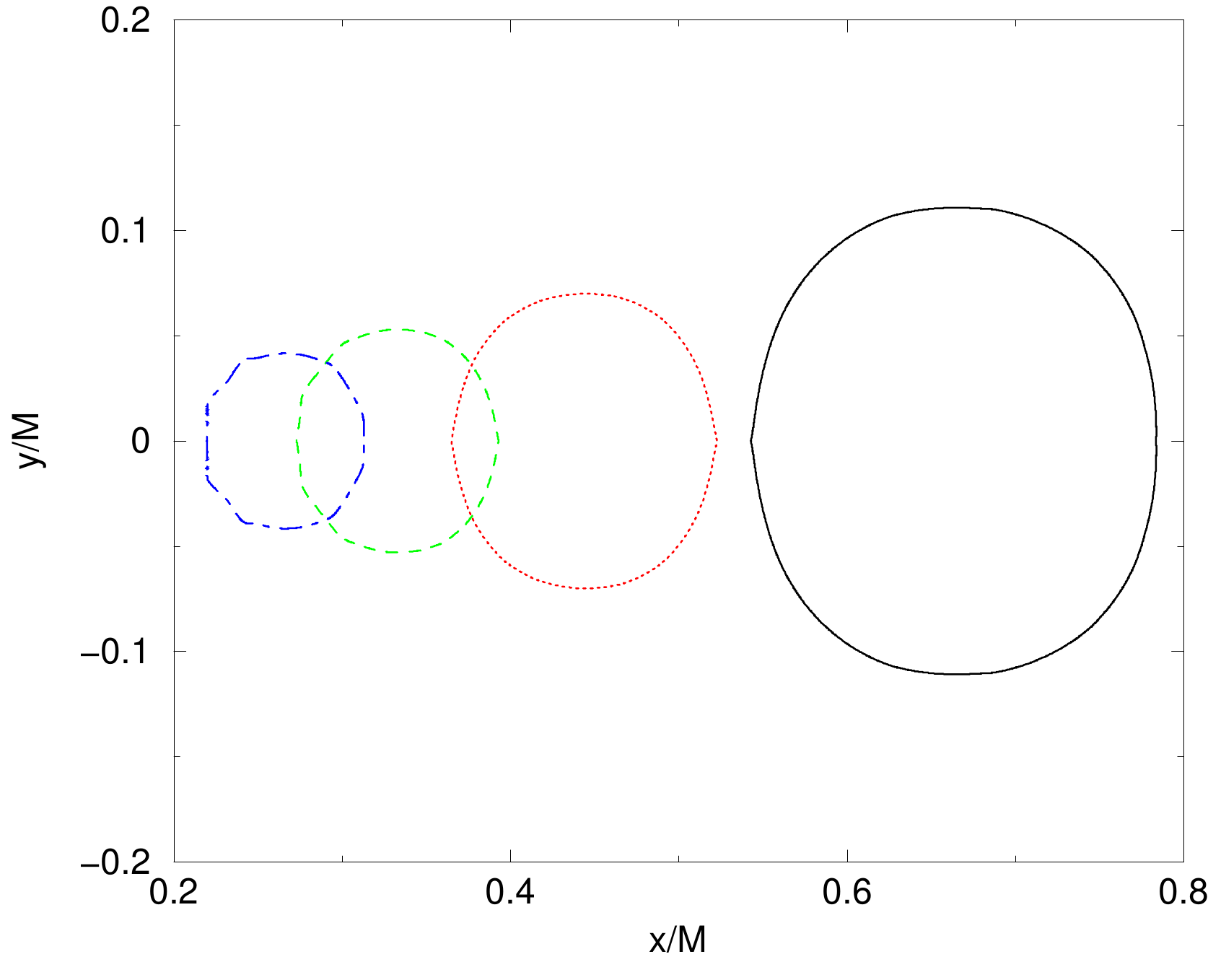}
  \includegraphics[width=.45\textwidth]{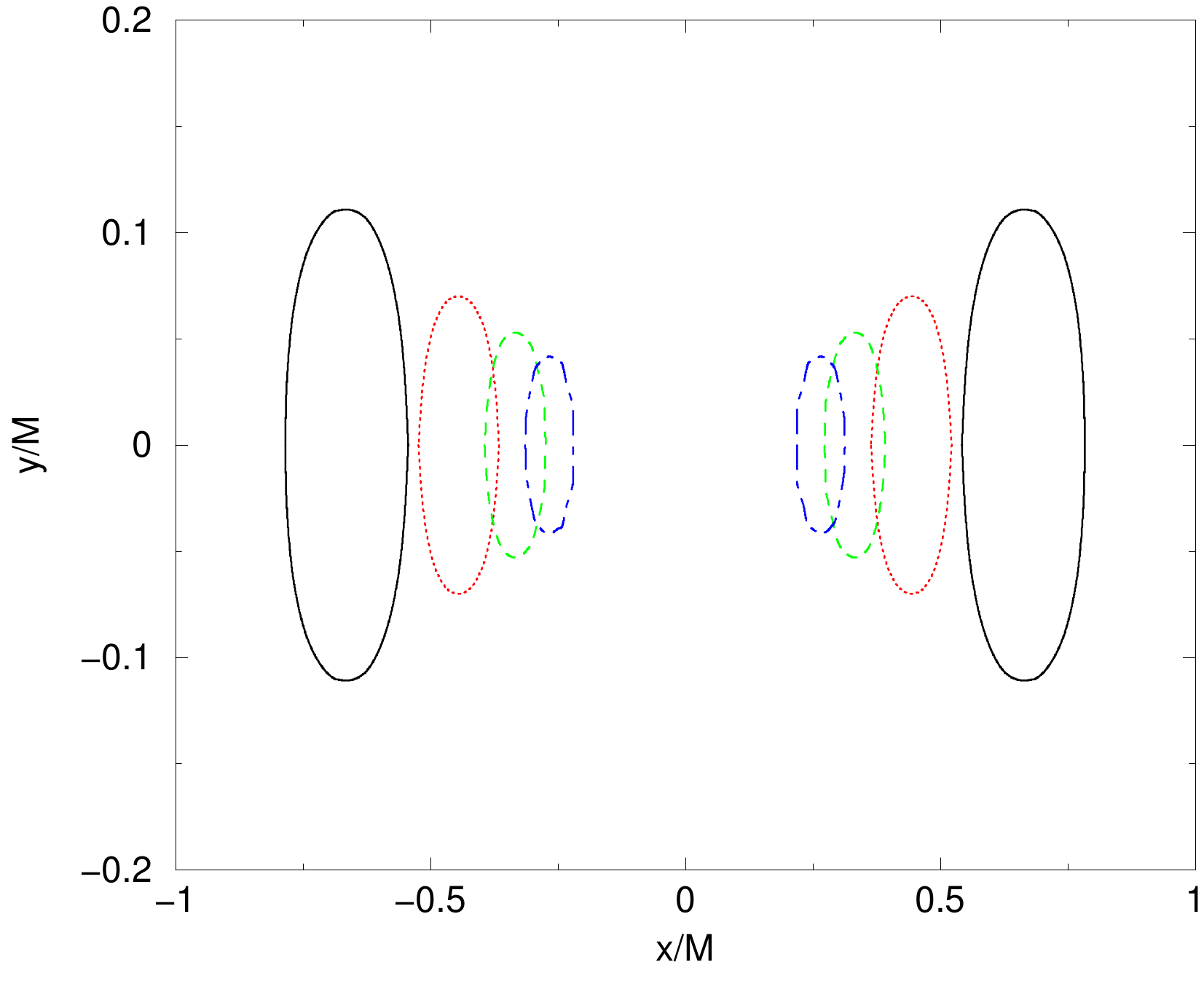}
  \end{center}
  \caption{The two innermost EHs near $t=0$ for the linear
distribution with $N=4$ (solid), $N=6$ (dotted), $N=8$ (dashed),
  and $N=10$ (dot-dashed) BHs. In this plot the coordinates have not
been rescaled. Note that as $N$ increases the two innermost horizons
approach each other, but also simultaneously shrink in size.
}
  \label{fig:line_BHs-noscaled}
\end{figure}

\begin{figure}
 \begin{center}
  \includegraphics[width=.45\textwidth]{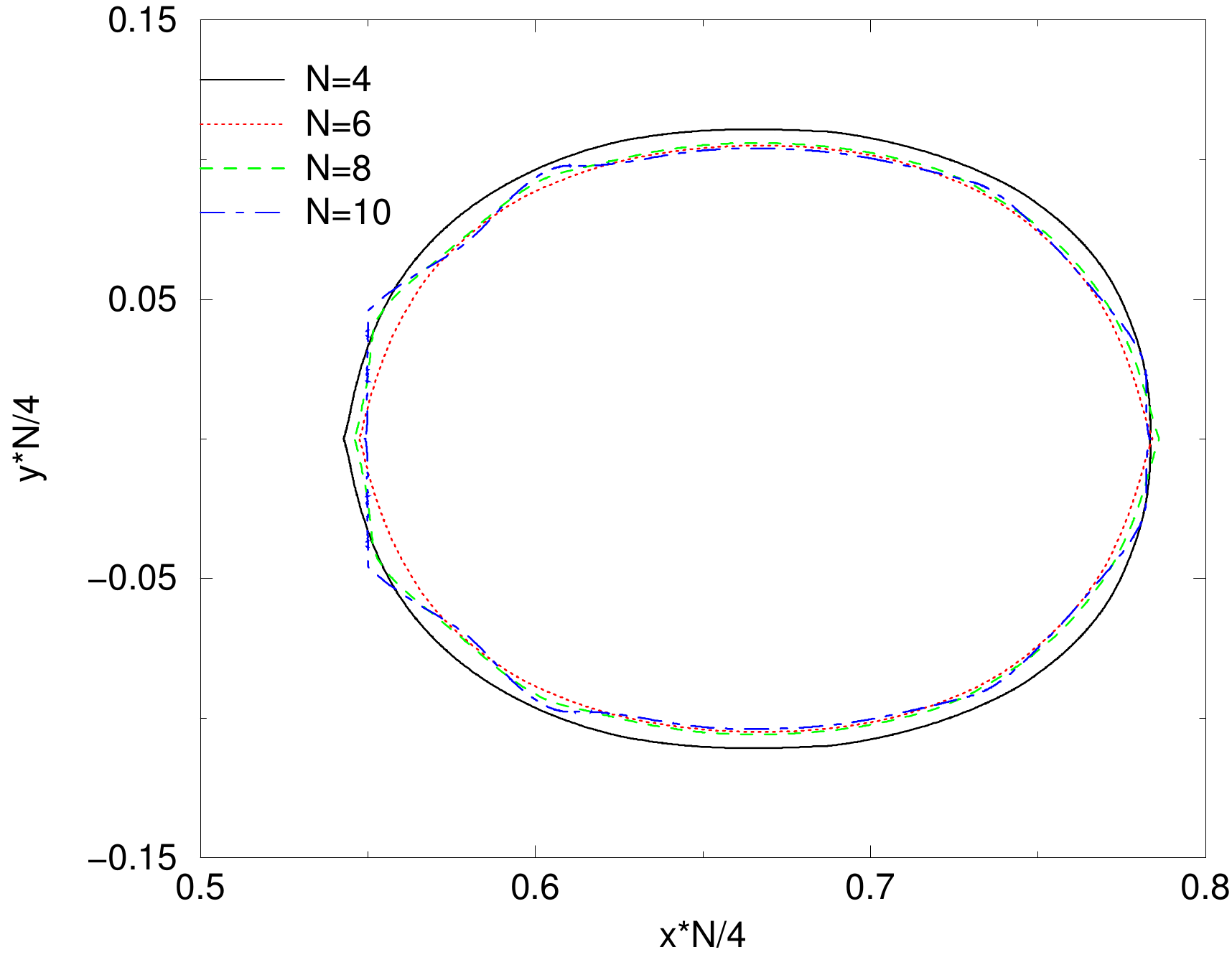}
  \includegraphics[width=.45\textwidth]{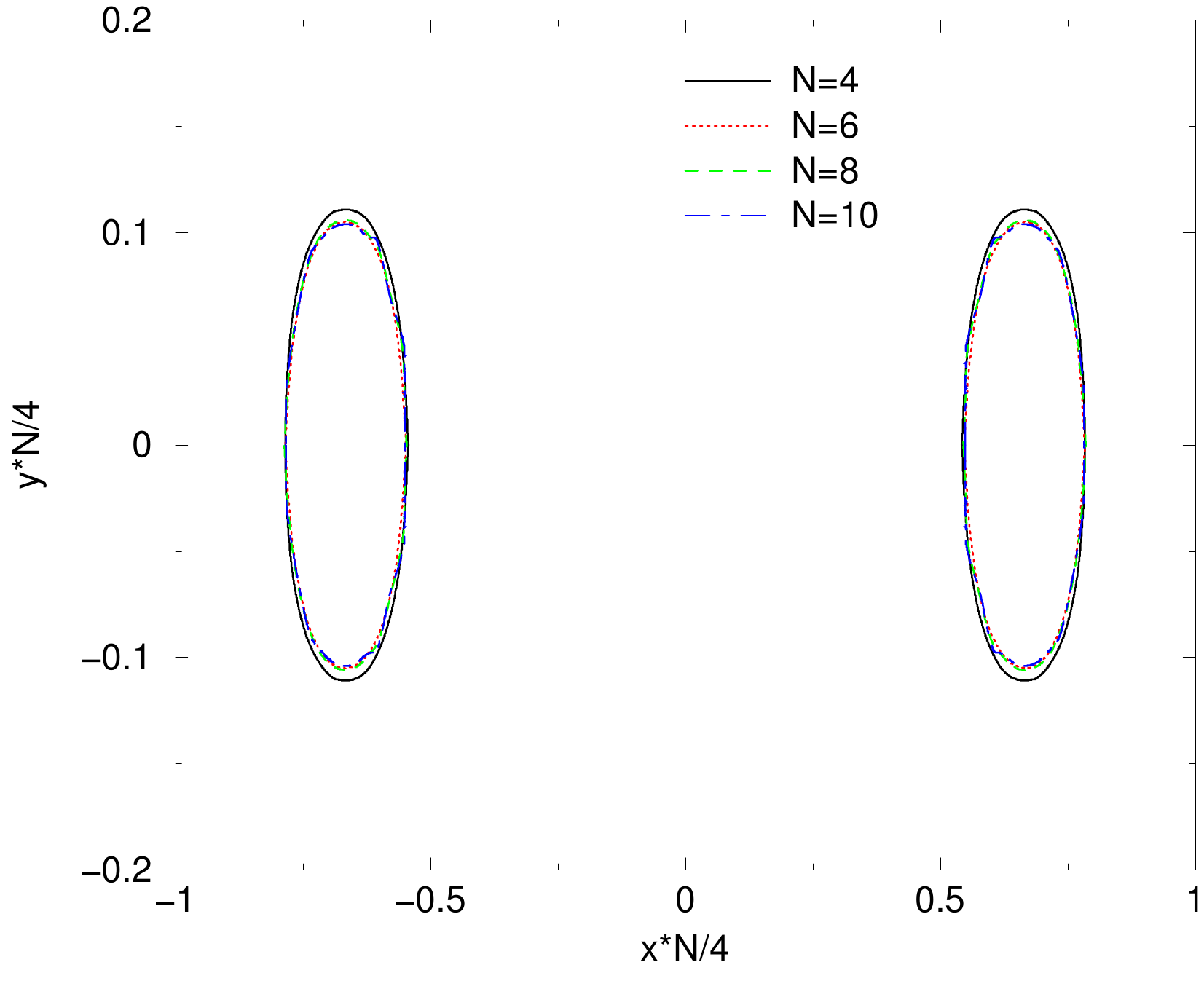}
\end{center}
  \caption{The two innermost EHs near $t=0$ for the linear
distribution with $N=4$ (solid), $N=6$ (dotted), $N=8$ (dashed),
  and $N=10$ (dot-dashed) BHs. Note that the distances have been
scaled by $N/4$ and that the innermost horizons do not appear to
approach each other as $N$ increases.}
  \label{fig:line_BHs_1st-scaled}
\end{figure}

In Figure~\ref{fig:line_allBHs-noscaled} we show the EHs at
$t\approx 0$ for the linear distribution with $N=4, 6, 8$, and $10$
BHs uniformly distributed over a line with length given by
Eq.~(\ref{eqn:line_length_N}), i.e.  lengths: $4.0, 4.444, 4.667,
4.8$. For these evolution we used resolutions of $M/16$, $M/32$,
$M/40$, and $M/50$ for $N=4$, $N=6$, $N=8$, and $N=10$, respectively.
Due to an instability in the EH
search, we could not find EHs at exactly $t=0$ in all cases.
In
Figs.~\ref{fig:line_BHs-noscaled}~and~\ref{fig:line_BHs_1st-scaled}
we show only the two innermost EHs.

  According to these figures, there is a clear trend towards reducing
the effective size of the event horizons while keeping their
separations, relative to the individual BH radius, fixed
when increasing $N$ (see for instance
Fig.~\ref{fig:line_BHs_1st-scaled}). This suggests that for finite
$N$ there will be $N$ distinct BHs rather than a common EH. Thus for
the limiting case $N \rightarrow \infty$, the event horizons would
seem to have
at most point like width, giving in the most optimistic scenario a
null-width connected ring.

This setup was examined using an axisymmetric code with up to
1000 BHs arranged on a line in Ref.~\cite{Abrahams:1992ru}. In that paper
it was found that no common AH exists for lines longer than $1.5M$.
Here we have extended this argument (somewhat) to no common event
horizon for lines larger than $4M$ (we did not do a systematic study
to determine the minimum line length for the existence of a common
event horizon as a function of the number of
BHs).

\section{Discussion}
  \label{sec:Discussion}

It is important to note that, unlike the case of discrete punctures,
where the puncture singularities are only coordinate
singularities, the singularities for these continuum linear
distributions are physical singularities. We can show this by noting
that the initial hypersurfaces are not geodesically complete. In the
vicinity of a linear distribution, the conformal factor $\psi$ will
behave like $\psi\sim \log \rho$, where $\rho$ is the coordinate
distance to the line. Consequently, the physical distance of a point
$\rho$ away from the line is $\delta s\sim \rho \log^2 \rho$, which is
finite. Next we note that the Kretschmann invariant
$K=R_{\alpha \beta \gamma \delta}R^{\alpha \beta \gamma \delta}$
 is singular and of the form $K\sim 1/[\rho^4 (\log \rho)^{10}]$.

We can strengthen the argument for a non-simply connected EH in the
continuum ring configuration (for large enough radius). We expect that
the ring singularity will undergo gravitational collapse to a point.
That is, the ring radius in quasi-isotropic coordinates will decrease
with time.  We therefore expect that the radius will get larger
the farther back we move in time (one could also consider a family
of initial data, where the mass of the ring is fixed to $1M$ but
the radius is made arbitrarily large).

\begin{figure}
  \begin{center}
  \includegraphics[width=.75\textwidth]{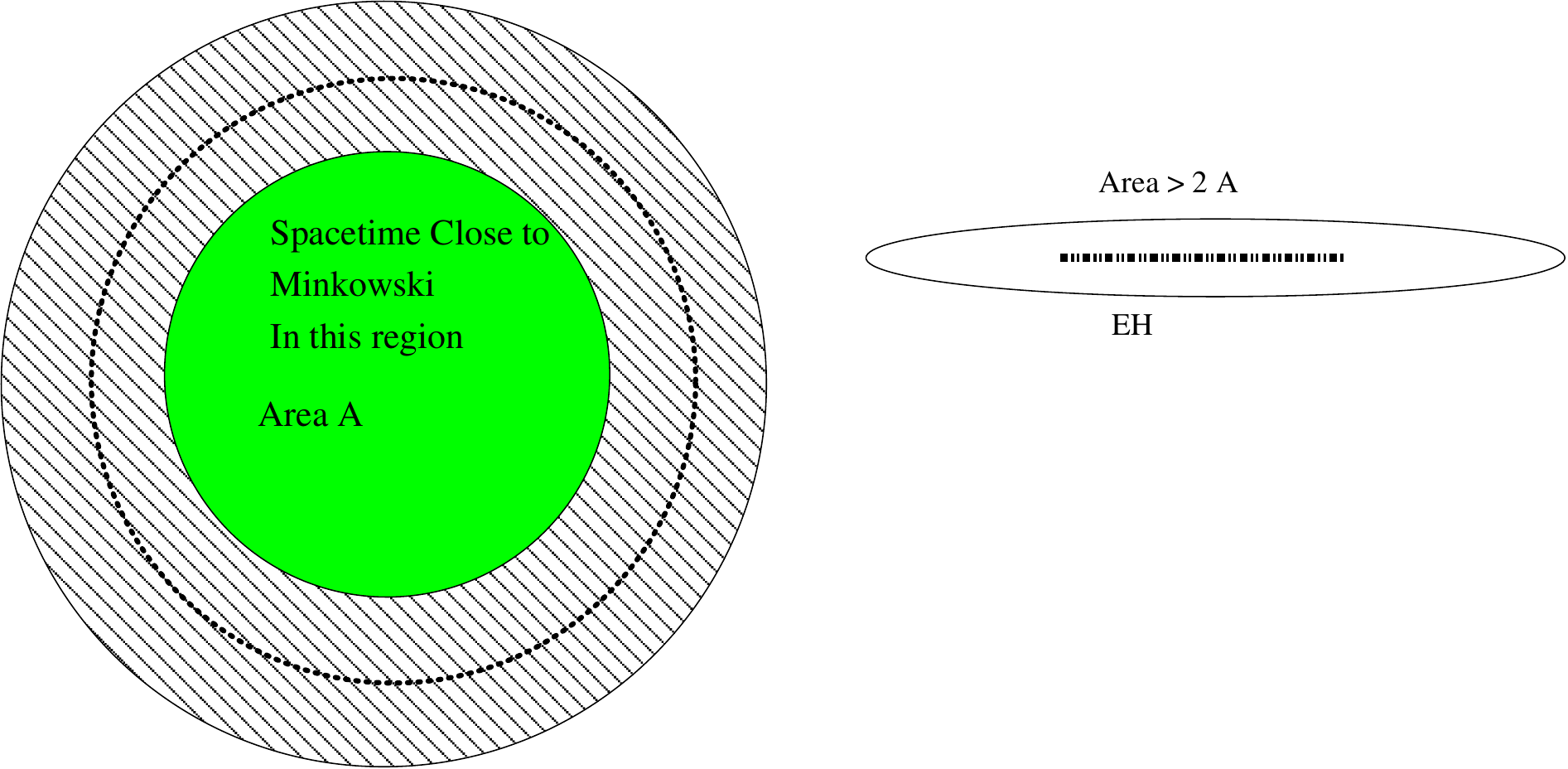}
  \end{center}
  \caption{The conjectured $S^2$ horizon outside the ring singularity
(dotted line) for a ring singularity with large radius. 
The shaded region is a region
inside the ring but still far enough from the singularity that the
metric is nearly flat. The surface area of this region is proportional
to the square of the ring's radius, and hence get arbitrarily large.
The horizon's area must be larger than twice this area. The plot on
the left shows a cut on the $xy$ plane, while the plot on the right
shows a cut on the $xz$ plane.}
  \label{fig:ring_area}
\end{figure}

Suppose
that there is an $S^2$ slice of ${\cal H}$ surrounding the ring at all
times. 
The surface area of the horizon
will become arbitrarily large (we note that metric in the neighborhood
of the ring is only logarithmically singular, and approaches Minkowski
rapidly with the distance from the ring. This means that
the
spacetime will be very nearly Minkowski, including in the center of
the ring) in the distant past (see Fig.~\ref{fig:ring_area}).
Thus if the ring is surrounded by an $S^2$ horizon 
in the distant past, 
the spacetime will be asymptotically predictable
(i.e.\ posses at least a partial Cauchy surface)  and should evolve to a
Schwarzschild BH with surface area at least as big as the surface of
the horizon when the ring is arbitrarily large (i.e. for the very
distant past). However, this
is not possible because the ring's mass is fixed, and hence its the
horizons area is bounded. Consequently, the EH, if it
exists, cannot by $S^2$. We note that our tests with discrete ring
configuration, and ring singularities with small radii, all indicate
that the horizon area increases with time (as required). The increase
in area is quite dramatic for the discrete ring. This is due to fact
that the configuration does not radiate significantly, so the
initial horizon masses are approximately equal to the total mass
divided by the number of black holes. Hence the area of each horizon
is proportional equal to $16 \pi (M/N)^2$ and the total initial area is
equal to $16 \pi M^2 / N$, or $1/N$ times the final horizon area. In
the limit $N\to\infty$ this leads to an initial horizon with vanishing
area.

We note that our results from the discrete ring configuration suggests
that the EH, if it exists, may have vanishing width.
If this is so, then null geodesics originating arbitrarily close to the
singularity (and hence in regions of arbitrarily large curvature) are
visible to null infinity $\scri^+$, indicating the presence of a
type of naked singularity, and the spacetime itself may not be
asymptotically predictable (the presence of a naked singularity does
not preclude that the spacetime is asymptotically predictable).
While a horizon ${\cal H}_t$ with a stable toroidal
topology does not appear to exist for the ring singularity, the
caustic structure of ${\cal H}$ indicates that instantaneous
 toroidal slices are possible.
Here the spacetime has no common EH at sufficiently early timeslices and only
first appears at a timeslice $\Sigma_h$ as an extended object with topology
$S^2$. However, because ${\cal H} \cap
\Sigma_h$ is an extended object, $\Sigma_h$ can be distorted to an alternate
slice $\tilde \Sigma_h$ such that $\tilde \Sigma_h \cap {\cal H}$ is
not simply connected (e.g.\ by distorting the timeslice such that the
center of the ring is slightly earlier in time than the neighboring
points). See Fig.~\ref{fig:Caustic}.

Although we started by looking for configurations with EHs with
toroidal slices, we actually have a more interesting case of a ring
singularity with an EH in the future of some slice
$\Sigma_0$ and a possible naked singularity in the past of $\Sigma_0$.
Although, in the slicing used here, the horizon appears to not have
toroidal topology, we note that the slicing can be distorted to
produce a toroidal horizon. This configuration 
appears to be a very interesting topic for further analytical 
investigations.

\ack
We thank J.Friedman, Badri Krishnan and Peter Diener for helpful discussions.
We thank the referees of this paper for their valuable advice and
suggested revisions.
We gratefully acknowledge NSF for financial support from grant
PHY-0722315, PHY-0653303, PHY-0714388, PHY-0722703, DMS-0820923,
PHY-0929114, and PHY-0969855; and NASA for financial support from grant NASA
07-ATFP07-0158 and HST-AR-11763.  Computational resources were
provided by Ranger cluster at TACC (Teragrid allocations TG-PHY080040N
and TG-PHY060027N) and by NewHorizons at RIT.

\section*{References}
\bibliographystyle{iopart-num}
\bibliography{../../Bibtex/references}

\providecommand{\newblock}{}
\begin{thebibliography}{10}
\expandafter\ifx\csname url\endcsname\relax
  \def\url#1{{\tt #1}}\fi
\expandafter\ifx\csname urlprefix\endcsname\relax\def\urlprefix{URL }\fi
\providecommand{\eprint}[2][]{\url{#2}}

\bibitem{Pretorius:2005gq}
Pretorius F 2005 {\em Phys. Rev. Lett.\/} {\bf 95} 121101 (\textit{Preprint}
  \eprint{gr-qc/0507014})

\bibitem{Campanelli:2005dd}
Campanelli M, Lousto C~O, Marronetti P and Zlochower Y 2006 {\em Phys. Rev.
  Lett.\/} {\bf 96} 111101 (\textit{Preprint} \eprint{gr-qc/0511048})

\bibitem{Baker:2005vv}
Baker J~G, Centrella J, Choi D~I, Koppitz M and van Meter J 2006 {\em Phys.
  Rev. Lett.\/} {\bf 96} 111102 (\textit{Preprint} \eprint{gr-qc/0511103})

\bibitem{Lindblom:2005qh}
Lindblom L, Scheel M~A, Kidder L~E, Owen R and Rinne O 2006 {\em Class. Quant.
  Grav.\/} {\bf 23} S447--S462 (\textit{Preprint} \eprint{gr-qc/0512093})

\bibitem{Gundlach:2006tw}
Gundlach C and Martin-Garcia J~M 2006 {\em Phys. Rev.\/} {\bf D74} 024016
  (\textit{Preprint} \eprint{gr-qc/0604035})

\bibitem{vanMeter:2006vi}
van Meter J~R, Baker J~G, Koppitz M and Choi D~I 2006 {\em Phys. Rev.\/} {\bf
  D73} 124011 (\textit{Preprint} \eprint{gr-qc/0605030})

\bibitem{Campanelli:2006uy}
Campanelli M, Lousto C~O and Zlochower Y 2006 {\em Phys. Rev. D\/} {\bf 74}
  041501(R) (\textit{Preprint} \eprint{gr-qc/0604012})

\bibitem{Campanelli:2006fg}
Campanelli M, Lousto C~O and Zlochower Y 2006 {\em Phys. Rev. D\/} {\bf 74}
  084023 (\textit{Preprint} \eprint{astro-ph/0608275})

\bibitem{Campanelli:2006fy}
Campanelli M, Lousto C~O, Zlochower Y, Krishnan B and Merritt D 2007 {\em Phys.
  Rev.\/} {\bf D75} 064030 (\textit{Preprint} \eprint{gr-qc/0612076})

\bibitem{Rezzolla:2007rd}
Rezzolla L {\em et~al.\/} 2008 {\em Astrophys. J.\/} {\bf 674} L29--L32
  (\textit{Preprint} \eprint{arXiv:0710.3345 [gr-qc]})

\bibitem{Sperhake:2007gu}
Sperhake U {\em et~al.\/} 2008 {\em Phys. Rev.\/} {\bf D78} 064069
  (\textit{Preprint} \eprint{0710.3823})

\bibitem{Dain:2008ck}
Dain S, Lousto C~O and Zlochower Y 2008 {\em Phys. Rev. D\/} {\bf 78} 024039
  (\textit{Preprint} \eprint{0803.0351})

\bibitem{Dreyer:2002mx}
Dreyer O, Krishnan B, Shoemaker D and Schnetter E 2003 {\em Phys. Rev.\/} {\bf
  D67} 024018 (\textit{Preprint} \eprint{gr-qc/0206008})

\bibitem{Schnetter:2006yt}
Schnetter E, Krishnan B and Beyer F 2006 {\em Phys. Rev.\/} {\bf D74} 024028
  (\textit{Preprint} \eprint{gr-qc/0604015})

\bibitem{Cook:2007wr}
Cook G~B and Whiting B~F 2007 {\em Phys. Rev.\/} {\bf D76} 041501
  (\textit{Preprint} \eprint{arXiv:0706.0199 [gr-qc]})

\bibitem{Krishnan:2007pu}
Krishnan B, Lousto C~O and Zlochower Y 2007 {\em Phys. Rev.\/} {\bf D76} 081501
  (\textit{Preprint} \eprint{0707.0876})

\bibitem{Campanelli:2008dv}
Campanelli M, Lousto C~O and Zlochower Y 2009 {\em Phys. Rev. D\/} {\bf 79}
  084012 (\textit{Preprint} \eprint{0811.3006})

\bibitem{Owen:2010vw}
Owen R 2010 {\em Phys. Rev.\/} {\bf D81} 124042 (\textit{Preprint}
  \eprint{1004.3768})

\bibitem{Herrmann:2007ex}
Herrmann F, Hinder I, Shoemaker D~M, Laguna P and Matzner R~A 2007 {\em Phys.
  Rev.\/} {\bf D76} 084032 (\textit{Preprint} \eprint{0706.2541})

\bibitem{Marronetti:2007ya}
Marronetti P {\em et~al.\/} 2007 {\em Class. Quant. Grav.\/} {\bf 24} S43--S58
  (\textit{Preprint} \eprint{gr-qc/0701123})

\bibitem{Marronetti:2007wz}
Marronetti P, Tichy W, Brugmann B, Gonzalez J and Sperhake U 2008 {\em Phys.
  Rev.\/} {\bf D77} 064010 (\textit{Preprint} \eprint{0709.2160})

\bibitem{Berti:2007fi}
Berti E {\em et~al.\/} 2007 {\em Phys. Rev.\/} {\bf D76} 064034
  (\textit{Preprint} \eprint{gr-qc/0703053})

\bibitem{Herrmann:2006ks}
Herrmann F, Shoemaker D and Laguna P 2006 {\em AIP Conf.\/} {\bf 873} 89--93
  (\textit{Preprint} \eprint{gr-qc/0601026})

\bibitem{Baker:2006vn}
Baker J~G {\em et~al.\/} 2006 {\em Astrophys. J.\/} {\bf 653} L93--L96
  (\textit{Preprint} \eprint{astro-ph/0603204})

\bibitem{Gonzalez:2006md}
Gonz\'alez J~A, Sperhake U, Brugmann B, Hannam M and Husa S 2007 {\em Phys.
  Rev. Lett.\/} {\bf 98} 091101 (\textit{Preprint} \eprint{gr-qc/0610154})

\bibitem{Herrmann:2007ac}
Herrmann F, Hinder I, Shoemaker D, Laguna P and Matzner R~A 2007 {\em
  Astrophys. J.\/} {\bf 661} 430--436 (\textit{Preprint}
  \eprint{gr-qc/0701143})

\bibitem{Campanelli:2007ew}
Campanelli M, Lousto C~O, Zlochower Y and Merritt D 2007 {\em Astrophys. J.\/}
  {\bf 659} L5--L8 (\textit{Preprint} \eprint{gr-qc/0701164})

\bibitem{Campanelli:2007cga}
Campanelli M, Lousto C~O, Zlochower Y and Merritt D 2007 {\em Phys. Rev.
  Lett.\/} {\bf 98} 231102 (\textit{Preprint} \eprint{gr-qc/0702133})

\bibitem{Lousto:2008dn}
Lousto C~O and Zlochower Y 2009 {\em Phys. Rev. D\/} {\bf 79} 064018
  (\textit{Preprint} \eprint{0805.0159})

\bibitem{Pollney:2007ss}
Pollney D {\em et~al.\/} 2007 {\em Phys. Rev.\/} {\bf D76} 124002
  (\textit{Preprint} \eprint{0707.2559})

\bibitem{Gonzalez:2007hi}
Gonz\'alez J~A, Hannam M~D, Sperhake U, Brugmann B and Husa S 2007 {\em Phys.
  Rev. Lett.\/} {\bf 98} 231101 (\textit{Preprint} \eprint{gr-qc/0702052})

\bibitem{Brugmann:2007zj}
Brugmann B, Gonzalez J~A, Hannam M, Husa S and Sperhake U 2008 {\em Phys.
  Rev.\/} {\bf D77} 124047 (\textit{Preprint} \eprint{0707.0135})

\bibitem{Choi:2007eu}
Choi D~I {\em et~al.\/} 2007 {\em Phys. Rev.\/} {\bf D76} 104026
  (\textit{Preprint} \eprint{gr-qc/0702016})

\bibitem{Baker:2007gi}
Baker J~G {\em et~al.\/} 2007 {\em Astrophys. J.\/} {\bf 668} 1140--1144
  (\textit{Preprint} \eprint{astro-ph/0702390})

\bibitem{Schnittman:2007ij}
Schnittman J~D {\em et~al.\/} 2008 {\em Phys. Rev.\/} {\bf D77} 044031
  (\textit{Preprint} \eprint{0707.0301})

\bibitem{Baker:2008md}
Baker J~G {\em et~al.\/} 2008 {\em Astrophys. J.\/} {\bf 682} L29
  (\textit{Preprint} \eprint{0802.0416})

\bibitem{Healy:2008js}
Healy J {\em et~al.\/} 2009 {\em Phys. Rev. Lett.\/} {\bf 102} 041101
  (\textit{Preprint} \eprint{0807.3292})

\bibitem{Herrmann:2007zz}
Herrmann F, Hinder I, Shoemaker D and Laguna P 2007 {\em Class. Quant. Grav.\/}
  {\bf 24} S33--S42

\bibitem{Tichy:2007hk}
Tichy W and Marronetti P 2007 {\em Phys. Rev.\/} {\bf D76} 061502
  (\textit{Preprint} \eprint{gr-qc/0703075})

\bibitem{Koppitz:2007ev}
Koppitz M {\em et~al.\/} 2007 {\em Phys. Rev. Lett.\/} {\bf 99} 041102
  (\textit{Preprint} \eprint{gr-qc/0701163})

\bibitem{Miller:2008en}
Miller S~H and Matzner R~A 2009 {\em Gen. Rel. Grav.\/} {\bf 41} 525--539
  (\textit{Preprint} \eprint{0807.3028})

\bibitem{Boyle:2007sz}
Boyle L, Kesden M and Nissanke S 2008 {\em Phys. Rev. Lett.\/} {\bf 100} 151101
  (\textit{Preprint} \eprint{0709.0299})

\bibitem{Boyle:2007ru}
Boyle L and Kesden M 2008 {\em Phys. Rev.\/} {\bf D78} 024017
  (\textit{Preprint} \eprint{0712.2819})

\bibitem{Buonanno:2007sv}
Buonanno A, Kidder L~E and Lehner L 2008 {\em Phys. Rev.\/} {\bf D77} 026004
  (\textit{Preprint} \eprint{arXiv:0709.3839 [astro-ph]})

\bibitem{Tichy:2008du}
Tichy W and Marronetti P 2008 {\em Phys. Rev.\/} {\bf D78} 081501
  (\textit{Preprint} \eprint{0807.2985})

\bibitem{Kesden:2008ga}
Kesden M 2008 {\em Phys. Rev.\/} {\bf D78} 084030 (\textit{Preprint}
  \eprint{0807.3043})

\bibitem{Barausse:2009uz}
Barausse E and Rezzolla L 2009 {\em Astrophys. J. Lett.\/} {\bf 704} L40--L44
  (\textit{Preprint} \eprint{0904.2577})

\bibitem{Rezzolla:2008sd}
Rezzolla L 2009 {\em Class. Quant. Grav.\/} {\bf 26} 094023 (\textit{Preprint}
  \eprint{0812.2325})

\bibitem{Lousto:2009mf}
Lousto C~O, Campanelli M, Zlochower Y and Nakano H 2010 {\em Class. Quant.
  Grav.\/} {\bf 27} 114006 (\textit{Preprint} \eprint{0904.3541})

\bibitem{Buonanno:2006ui}
Buonanno A, Cook G~B and Pretorius F 2007 {\em Phys. Rev.\/} {\bf D75} 124018
  (\textit{Preprint} \eprint{gr-qc/0610122})

\bibitem{Baker:2006ha}
Baker J~G, van Meter J~R, McWilliams S~T, Centrella J and Kelly B~J 2007 {\em
  Phys. Rev. Lett.\/} {\bf 99} 181101 (\textit{Preprint}
  \eprint{gr-qc/0612024})

\bibitem{Pan:2007nw}
Pan Y {\em et~al.\/} 2008 {\em Phys. Rev.\/} {\bf D77} 024014
  (\textit{Preprint} \eprint{arXiv:0704.1964 [gr-qc]})

\bibitem{Buonanno:2007pf}
Buonanno A {\em et~al.\/} 2007 {\em Phys. Rev.\/} {\bf D76} 104049
  (\textit{Preprint} \eprint{arXiv:0706.3732 [gr-qc]})

\bibitem{Hannam:2007ik}
Hannam M, Husa S, Sperhake U, Brugmann B and Gonzalez J~A 2008 {\em Phys.
  Rev.\/} {\bf D77} 044020 (\textit{Preprint} \eprint{0706.1305})

\bibitem{Hannam:2007wf}
Hannam M, Husa S, Bruegmann B and Gopakumar A 2008 {\em Phys. Rev.\/} {\bf D78}
  104007 (\textit{Preprint} \eprint{0712.3787})

\bibitem{Gopakumar:2007vh}
Gopakumar A, Hannam M, Husa S and Bruegmann B 2008 {\em Phys. Rev.\/} {\bf D78}
  064026 (\textit{Preprint} \eprint{0712.3737})

\bibitem{Hinder:2008kv}
Hinder I, Herrmann F, Laguna P and Shoemaker D 2010 {\em Phys. Rev.\/} {\bf
  D82} 024033 (\textit{Preprint} \eprint{0806.1037})

\bibitem{Gonzalez:2008bi}
Gonzalez J~A, Sperhake U and Brugmann B 2009 {\em Phys. Rev.\/} {\bf D79}
  124006 (\textit{Preprint} \eprint{0811.3952})

\bibitem{Lousto:2010tb}
Lousto C~O, Nakano H, Zlochower Y and Campanelli M 2010 {\em Phys. Rev.
  Lett.\/} {\bf 104} 211101 (\textit{Preprint} \eprint{1001.2316})

\bibitem{Friedman:1993ty}
Friedman J~L, Schleich K and Witt D~M 1993 {\em Phys. Rev. Lett.\/} {\bf 71}
  1486--1489 (\textit{Preprint} \eprint{gr-qc/9305017})

\bibitem{Hawking:1971vc}
Hawking S~W 1972 {\em Commun. Math. Phys.\/} {\bf 25} 152--166

\bibitem{Hawking73a}
Hawking S~W and Ellis G~F~R 1973 {\em The Large Scale Structure of Spacetime\/}
  (Cambridge, England: Cambridge University Press)

\bibitem{Galloway:2005mf}
Galloway G~J and Schoen R 2006 {\em Commun. Math. Phys.\/} {\bf 266} 571--576
  (\textit{Preprint} \eprint{gr-qc/0509107})

\bibitem{Galloway:2006ws}
Galloway G~J 2006  (\textit{Preprint} \eprint{gr-qc/0608118})

\bibitem{Racz:2008tf}
Racz I 2008 {\em Class. Quant. Grav.\/} {\bf 25} 162001 (\textit{Preprint}
  \eprint{0806.4373})

\bibitem{Hughes:1994ea}
Hughes S~A {\em et~al.\/} 1994 {\em Phys. Rev.\/} {\bf D49} 4004--4015

\bibitem{Shapiro:1995rr}
Shapiro S~L, Teukolsky S~A and Winicour J 1995 {\em Phys. Rev.\/} {\bf D52}
  6982--6987

\bibitem{Husa:1999nm}
Husa S and Winicour J 1999 {\em Phys. Rev.\/} {\bf D60} 084019
  (\textit{Preprint} \eprint{gr-qc/9905039})

\bibitem{Diener:2003jc}
Diener P 2003 {\em Class. Quant. Grav.\/} {\bf 20} 4901--4918
  (\textit{Preprint} \eprint{gr-qc/0305039})

\bibitem{Cohen:2008wa}
Cohen M~I, Pfeiffer H~P and Scheel M~A 2009 {\em Class. Quant. Grav.\/} {\bf
  26} 035005 (\textit{Preprint} \eprint{0809.2628})

\bibitem{Shapiro:1991zza}
Shapiro S~L and Teukolsky S~A 1991 {\em Phys. Rev. Lett.\/} {\bf 66} 994--997

\bibitem{PhysRevD.45.2006}
Shapiro S~L and Teukolsky S~A 1992 {\em Phys. Rev. D\/} {\bf 45} 2006--2012

\bibitem{Abrahams:1992ru}
Abrahams A~M, Heiderich K~R, Shapiro S~L and Teukolsky S~A 1992 {\em Phys.
  Rev.\/} {\bf D46} 2452--2463

\bibitem{Lehner:2010pn}
Lehner L and Pretorius F 2010 {\em Phys. Rev. Lett.\/} {\bf 105} 101102
  (\textit{Preprint} \eprint{1006.5960})

\bibitem{Jaramillo:2010tc}
Jaramillo G and Lousto C~O 2010  (\textit{Preprint} \eprint{1008.2001})

\bibitem{PhysRevD.41.1867}
Wojtkiewicz J 1990 {\em Phys. Rev. D\/} {\bf 41} 1867--1874

\bibitem{Brill63}
{B}rill D and Lindquist R 1963 {\em Phys. Rev.\/} {\bf 131} 471--476

\bibitem{Zlochower:2005bj}
Zlochower Y, Baker J~G, Campanelli M and Lousto C~O 2005 {\em Phys. Rev. D\/}
  {\bf 72} 024021 (\textit{Preprint} \eprint{gr-qc/0505055})

\bibitem{Alcubierre02a}
Alcubierre M, Br\"ugmann B, Diener P, Koppitz M, Pollney D, Seidel E and
  Takahashi R 2003 {\em Phys. Rev. D\/} {\bf 67} 084023 (\textit{Preprint}
  \eprint{gr-qc/0206072})

\bibitem{cactus_web}
Cactus Computational Toolkit home page: {\tt http://www.cactuscode.org}

\bibitem{Schnetter-etal-03b}
Schnetter E, Hawley S~H and Hawke I 2004 {\em Class. Quantum Grav.\/} {\bf 21}
  1465--1488 (\textit{Preprint} \eprint{gr-qc/0310042})

\bibitem{Thornburg2003:AH-finding}
Thornburg J 2004 {\em Class. Quantum Grav.\/} {\bf 21} 743--766
  (\textit{Preprint} \eprint{gr-qc/0306056})

\bibitem{Baker:2001sf}
Baker J, Campanelli M and Lousto C~O 2002 {\em Phys. Rev. D\/} {\bf 65} 044001
  (\textit{Preprint} \eprint{gr-qc/0104063})

\bibitem{Campanelli:2007ea}
Campanelli M, Lousto C~O and Zlochower Y 2008 {\em Phys. Rev. D\/} {\bf 77}
  101501(R) (\textit{Preprint} \eprint{0710.0879})

\bibitem{Lousto:2007rj}
Lousto C~O and Zlochower Y 2008 {\em Phys. Rev.\/} {\bf D77} 024034
  (\textit{Preprint} \eprint{0711.1165})

\bibitem{Galaviz:2010mx}
Galaviz P, Brugmann B and Cao Z 2010 {\em Phys. Rev.\/} {\bf D82} 024005
  (\textit{Preprint} \eprint{1004.1353})

\end{thebibliography}

\end{document}